\documentclass[times, trackchanges]{aastex7}
\usepackage{amsmath}
\usepackage{enumitem}
\usepackage{hyperref}
\usepackage{physics}

\begin{document}
\title{Unification Model of Active Galactic Nuclei by \\ Photoionization Equilibrium Calculation Based on Radiative Hydrodynamic Simulations}
\author[0000-0002-0114-5581]{Atsushi Tanimoto}
\affiliation{Graduate School of Science and Engineering, Kagoshima University, Kagoshima 890-0065, Japan}
\email{atsushi.tanimoto@sci.kagoshima-u.ac.jp}
\author[0000-0002-8779-8486]{Keiichi Wada}
\affiliation{Graduate School of Science and Engineering, Kagoshima University, Kagoshima 890-0065, Japan}
\affiliation{Research Center for Space and Cosmic Evolution, Ehime University, Matsuyama 790-8577, Japan}
\affiliation{Faculty of Science, Hokkaido University, Sapporo 060-0810, Japan}
\email{wada@astrophysics.jp}
\author[0000-0003-0548-1766]{Yuki Kudoh}
\affiliation{Astronomical Institute, Tohoku University, Miyagi 980-8578, Japan}
\email{yuki.kudoh@astr.tohoku.ac.jp}
\author[0000-0003-2535-5513]{Nozomu Kawakatu}
\affiliation{Faculty of Natural Sciences, National Institute of Technology, Kure College, Hiroshima 737-8506, Japan}
\email{kawakatsu@kure-nct.ac.jp}
\author[0000-0002-6236-5270]{Mariko Nomura}
\affiliation{Graduate School of Science and Technology, Hirosaki University, Aomori, 036-8561, Japan}
\email{nomura@hirosaki-u.ac.jp}
\author[0000-0003-2670-6936]{Hirokazu Odaka}
\affiliation{Department of Earth and Space Science, The University of Osaka, Osaka 560-0043, Japan}
\email{odaka@ess.sci.osaka-u.ac.jp}

\begin{abstract}
To investigate the origin of the dependence of the covering factor on the Eddington ratio suggested by X-ray observations, we examined the angular distribution of \ion{H}{1} and \ion{H}{2} based on two-dimensional radiative hydrodynamic simulations. To calculate the Compton-thin covering factor $C_{22}$ and Compton-thick covering factor $C_{24}$ of \ion{H}{1} alone, we performed one-dimensional photoionization equilibrium calculations with the XSTAR code based on radiative hydrodynamic simulations. The results obtained are as follows. (1) The Compton-thin covering factor $C_{22}$ of \ion{H}{1} and \ion{H}{2} is independent of the Eddington ratio and is approximately $70\%$, while $C_{22}$ of \ion{H}{1} alone is also independent of the Eddington ratio and is approximately $30\%$. (2) The Compton-thick covering factor $C_{24}$ of \ion{H}{1} has the same value as $C_{22}$ of \ion{H}{1}. (3) Our $C_{24}$ is consistent with that obtained from X-ray observations. (4) Our $C_{22}$ agrees with that obtained from X-ray observations in a high Eddington ratio, while our $C_{22}$ is smaller than that from X-ray observations in a low Eddington ratio. (5) To explain the difference between $C_{22}$ obtained from theoretical calculations and that inferred from X-ray observations, a Compton-thin gas is required in regions extending at least $10~\mathrm{pc}$ beyond the current computational regions.
\end{abstract}
\section{Introduction}\label{Section0100}
The unified model of the active galactic nucleus (AGN) posits that a torus composed of gas and dust exists around an accreting supermassive black hole (SMBH) \citep{Antonucci1993, Urry1995, Netzer2015, Ramos2017}. The ratio of the solid angle covered by this torus to the SMBH is called the covering factor. Since this covering factor determines the ratio of unobscured to obscured AGNs, estimating its value is crucial. However, the formation mechanisms of the AGN torus and the precise value of the covering factor remain poorly understood.

One of the best tools for estimating the covering factor is X-ray spectroscopic observations. The X-ray spectrum of AGNs consists mainly of two components: a transmitted component absorbed by the torus through photoelectric absorption and a scattered component scattered by the torus through Compton scattering. This allows us to measure the hydrogen column density along the line of sight $N_{\mathrm{H}}^{\mathrm{LOS}}$ based on the shape of the transmitted component \citep[e.g.,][]{Ueda2003, Ueda2014, Kawamuro2016, Tanimoto2016, Tanimoto2018, Tanimoto2020, Ricci2017b}. In the case of X-ray spectroscopic observations, we classify sources with $N_{\mathrm{H}}^{\mathrm{LOS}}$ smaller than $10^{22} \ \mathrm{cm}^{-2}$ as unobscured AGNs, those with $N_{\mathrm{H}}^{\mathrm{LOS}}$ between $10^{22} \ \mathrm{cm}^{-2}$ and $10^{24} \ \mathrm{cm}^{-2}$ as Compton-thin AGNs, and those with $N_{\mathrm{H}}^{\mathrm{LOS}}$ greater than $10^{24} \ \mathrm{cm}^{-2}$ as Compton-thick AGNs \citep[e.g.,][]{Hickox2018}. That is, two types of covering factors can be defined. Here we introduce the Compton-thin covering factor $C_{22}$ and the Compton-thick covering factor $C_{24}$. The Compton-thin covering factor $C_{22}$ is the fraction of the central source covered by gas with $N_{\mathrm{H}}^{\mathrm{LOS}}$ greater than $10^{22} \ \mathrm{cm}^{-2}$ and the Compton-thick covering factor $C_{24}$ is the fraction greater than $10^{24} \ \mathrm{cm}^{-2}$. By measuring the hydrogen column density along the line of sight for a large number of sources, we can statistically determine $C_{22}$ and $C_{24}$.

\cite{Ricci2017b} systematically analyzed the X-ray spectra of $800$ AGN detected by the all-sky hard X-ray survey of the Neil Gehrels Swift Observatory \citep{Gehrels2004, Markwardt2005, Tueller2010, Baumgartner2013, Oh2018} and \cite{Ricci2017a} suggested the radiation-regulated AGN unification model. In this model, the Eddington ratio is a key parameter in determining $C_{22}$. If the Eddington ratio is less than $10^{-2}$, $C_{22}$ is a large value of $80\%$--$90\%$. On the other hand, when the Eddington ratio is greater than $10^{-2}$, $C_{22}$ is a small value of $30\%$--$40\%$. This is because radiation pressure is considered to blow away the dusty gas \cite[][Figure~4]{Ricci2017a}. However, in the X-ray spectral analysis of unobscured AGNs and Compton-thin AGNs, they assumed a simple Compton scattering model (pexrav; \citealt{Magdziarz1995}), which is based on flat-plate geometry. This assumption may be inconsistent with their high derived covering factor of $80\%$--$90\%$.

One of the leading candidates for the AGN torus formation mechanism is the radiation-driven fountain model \citep{Wada2012}. In the radiation-driven fountain model, a geometrically thick torus is naturally formed by the anisotropic radiation pressure on the dusty gas and the X-ray heating. In such theoretical calculations, the three-dimensional gas density distribution can be obtained, which allows the covering factor to be calculated directly. The model has successfully reproduced radio \citep{Wada2018a, Izumi2018, Izumi2023, Uzuo2021, Baba2024}, infrared \citep{Wada2016, Matsumoto2022, Matsumoto2023}, optical \citep{Wada2018b, Wada2023}, and X-ray observations \citep{Ogawa2022, Tanimoto2023, Tanimoto2025}. These were mainly compared with observations of the Circinus Galaxy, and an Eddington ratio of $20\%$ was assumed.

\cite{Kudoh2024} performed two-dimensional axisymmetric radiative hydrodynamic simulations for four Eddington ratios of $10^{-3}$, $10^{-2}$, $10^{-1}$, and $10^{0}$ with coordinated astronomical numerical software plus (CANS+: \citealt{Matsumoto2019}). \cite{Kudoh2024} calculated $C_{22}$ for each Eddington ratio based on their radiative hydrodynamic simulations and compared it to the radiation-regulated AGN unification model. As a result, their $C_{22}$ agrees with that inferred from X-ray observations in the low Eddington ratio, whereas their $C_{22}$ is larger than that inferred from X-ray observations in the high Eddington ratio \citep[][Figure~9]{Kudoh2024}. However, their calculation did not take into account whether hydrogen was \ion{H}{1} or \ion{H}{2}. Since \ion{H}{2} has no electrons, it does not absorb X-rays due to photoelectric absorption. To obtain the covering factor from theoretical calculations and compare it with that obtained from X-ray observations, it is necessary to calculate $C_{22}$ and $C_{24}$ for neutral material.

In this study, to calculate the covering factor of \ion{H}{1}, we performed photoionization equilibrium calculations with the XSTAR code \citep{Kallman2004} based on two-dimensional axisymmetric radiative hydrodynamic simulations for four Eddington ratios. The remainder of this paper is organized as follows. \hyperref[Section0200]{Section~2} presents the radiative hydrodynamic simulations and the photoionization equilibrium calculations. \hyperref[Section0300]{Section~3} describes the results of the radiative hydrodynamic simulations and the photoionization equilibrium calculations. \hyperref[Section0400]{Section~4} compares the covering factor obtained from our calculations with that inferred from X-ray observations. Finally, \hyperref[Section0500]{Section~5} presents our conclusions.\clearpage
\begin{figure*}\label{Figure0001}
\gridline{\fig{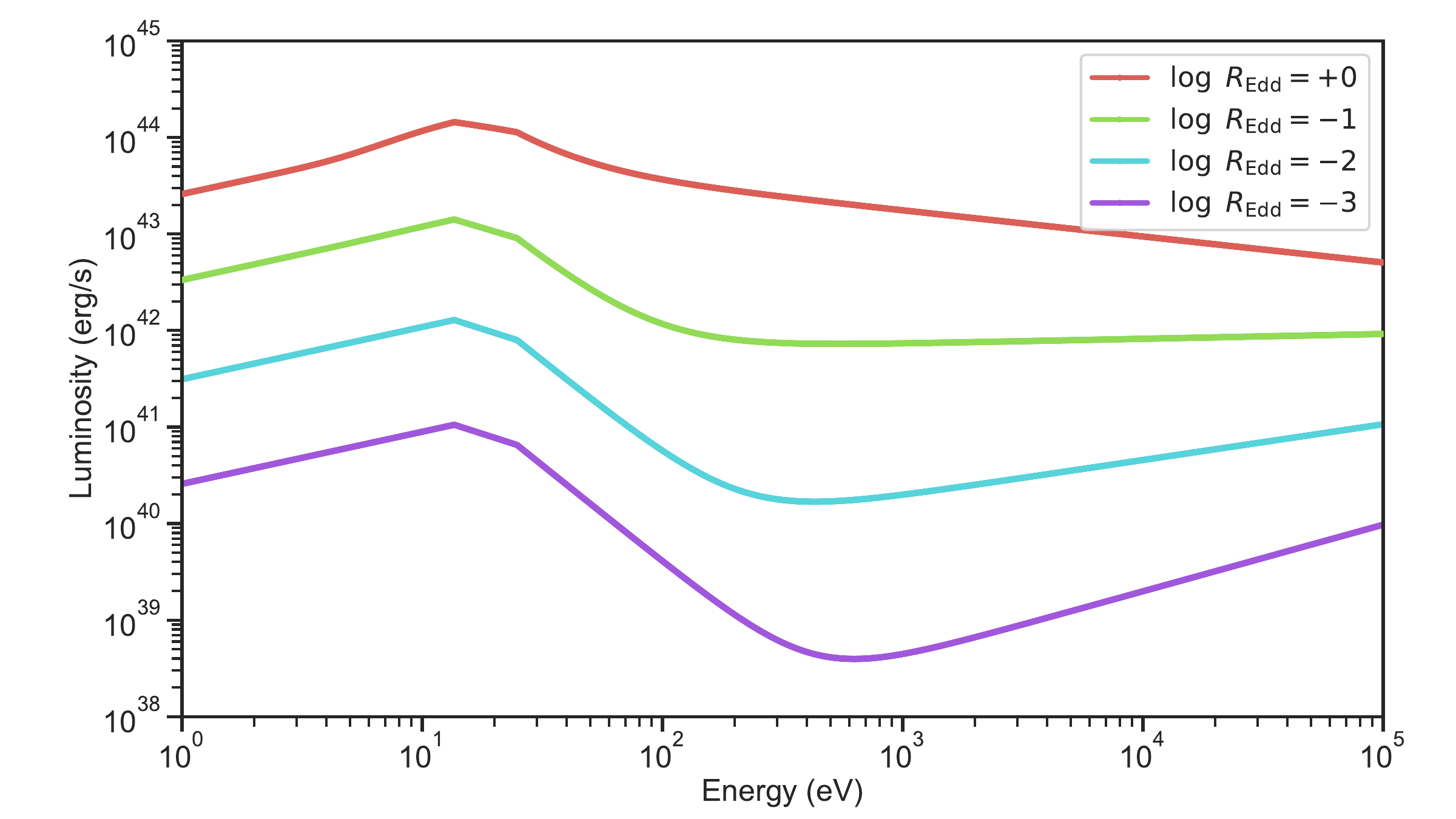}{0.5\textwidth}{}}
\caption{The spectral energy distribution (SED). The red, green, light blue, and purple lines correspond to the SED of the logarithmic Eddington ratio $\log R_{\mathrm{Edd}} = 0,-1,-2,\ \mathrm{and}\ -3$, respectively.}
\end{figure*}
\section{Methods}\label{Section0200}
To investigate the \ion{H}{1} covering factor, we perform photoionization equilibrium calculations based on two-dimensional axisymmetric radiative hydrodynamic simulations. \hyperref[Section0201]{Section~2.1} describes two-dimensional axisymmetric radiative hydrodynamic simulations and \hyperref[Section0202]{Section~2.2} presents one-dimensional photoionized equilibrium calculations.
\subsection{Radiative Hydrodynamic Simulations}\label{Section0201}
\cite{Kudoh2024} performed two-dimensional axisymmetric radiative hydrodynamic simulations of the dusty gas with coordinated astronomical numerical software plus (CANS+: \citealt{Matsumoto2019}). They solved the following basic equations numerically.
\begin{align}
\frac{\partial \rho}{\partial t} + \nabla \cdot (\rho \vb{v})                                           & = 0                         \\
\frac{\partial}{\partial t}(\rho \vb{v}) + \nabla \cdot (\rho \vb{v} \vb{v} + P_{\mathrm{gas}} \vb{I})  & = \vb{F}_{\mathrm{grav}} + \vb{F}_{\mathrm{radi}} + \vb{F}_{\mathrm{vis}}                                                                                                                 \\
\frac{\partial e}{\partial t} + \nabla \cdot (e\vb{v} + P_{\mathrm{gas}}\vb{v})                         & = \vb{v} \cdot \vb{F}_{\mathrm{grav}} + \vb{v} \cdot \vb{F}_{\mathrm{radi}} + W_{\mathrm{vis}} - \rho \mathcal{L}
\end{align}
where $\rho$ is the total density of gas and dust, $\vb{v}$ is the velocity, $P_{\mathrm{gas}}$ is the gas pressure, $\vb{F}_{\mathrm{grav}}$ is the gravitational force, $\vb{F}_{\mathrm{radi}}$ is the radiation force, $\vb{F}_{\mathrm{vis}}$ is the viscous force, $e$ is the total energy density, $W_{\mathrm{vis}}$ is the viscous heating and $\mathcal{L}$ is the heating/cooling rate per unit mass. They assumed that the mass of SMBH was $10^{7} M_{\sun}$ and the dust-to-gas ratio is $1\%$.

They performed two-dimensional axisymmetric radiative hydrodynamic simulations for four different Eddington ratios of $\log R_{\mathrm{Edd}} = -3, -2, -1, \ \mathrm{and} \ 0$. \hyperref[Figure0001]{Figure~1} shows the spectral energy density (SED). The SED is composed of an accretion disk component and a corona component. The component of the accretion disk is based on the Shakura-Sunyaev disk \citep{Shakura1973, Schartmann2005, Schartmann2011} and the component of the corona is assumed to be the power law $L_{\mathrm{corona}}(E) \propto E^{-\Gamma+1}$. In their study, the photon index $\Gamma$ depends on the Eddington ratio as $\Gamma = 0.32 \log R_{\mathrm{Edd}} + 2.27$ \citep{Brightman2013}. The computational domains in the cylindrical coordinate system are $10^{-4} \ \mathrm{pc} < r < 2 \ \mathrm{pc}$ and $-2 \ \mathrm{pc} < z < +2 \ \mathrm{pc}$. The boundary condition in the central region of $r \leq 2 \times 10^{-3} \ \mathrm{pc}$ is the absorbed boundary conditions with $\rho = 10^{-28} \ \mathrm{g} \ \mathrm{cm}^{-3}$ and $T_{\mathrm{gas}} = 10^{4} \ \mathrm{K}$. The boundary condition in the outer region is the boundary conditions of the outflow.\clearpage
\subsection{Photoionization Equilibrium Calculations}\label{Section0202}
We perform photoionization equilibrium calculations using the XSTAR code \citep{Kallman2004} based on two-dimensional axisymmetric radiative hydrodynamic simulations \citep{Kudoh2024}. The XSTAR is a code for calculating the physical conditions and spectra of ionized and nearly neutral gasses. Since XSTAR is a one-zone photoionization equilibrium calculation code, we performed XSTAR calculations for each cell and calculated the photoionization equilibrium from the inside to the outside.

The following five parameters are required to perform XSTAR calculations. (1) inner radius, (2) outer radius, (3) hydrogen number density in the region, (4) shape of the X-ray spectrum and (5) X-ray luminosity at $1$--$1000 \ \mathrm{Ry}$. (1)--(2) We generate a total of $9,000$ cells, with the $0.0 \ \mathrm{pc}$ to $0.1 \ \mathrm{pc}$ region divided $100$ in the radial direction and the $0\degr$ to $90\degr$ region divided $90$ in the angular direction. Although \cite{Kudoh2024} calculated density distributions up to $2~\mathrm{pc}$ (\hyperref[Section0201]{Section~2.1}), the inner region has a higher number density. Hence, we performed photoionization equilibrium calculations for the $0.0$--$0.1~\mathrm{pc}$ region. We confirmed that even when considering up to $2~\mathrm{pc}$, it does not affect the results obtained. (3) We assumed the hydrogen number density obtained from the radiative hydrodynamic simulations. (4)--(5) In the first cell, the same SED as in the radiative hydrodynamic simulation is assumed. For the second and subsequent cells, the luminosity and spectrum output of the photoionization equilibrium calculations of the previous cell were used. We note that although \cite{Kudoh2024} considered the angular dependence of the disk component, we assumed that the disk component is isotropic. This is because the ionization state is primarily determined by the corona component in our calculations. The solar abundances of \cite{Anders1989} were assumed.\clearpage
\begin{figure*}\label{Figure0002}
\gridline{\fig{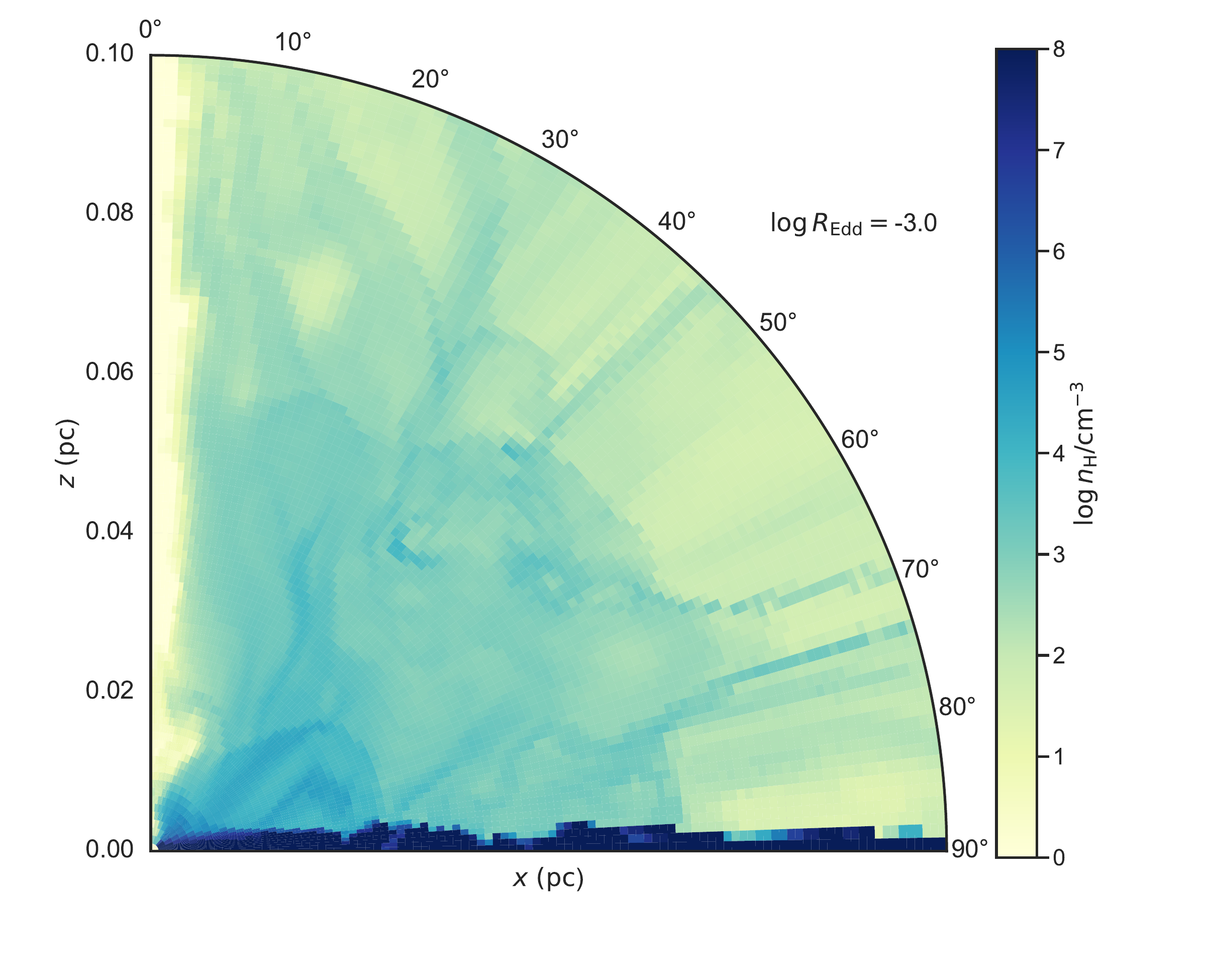}{0.5\textwidth}{(a)}\fig{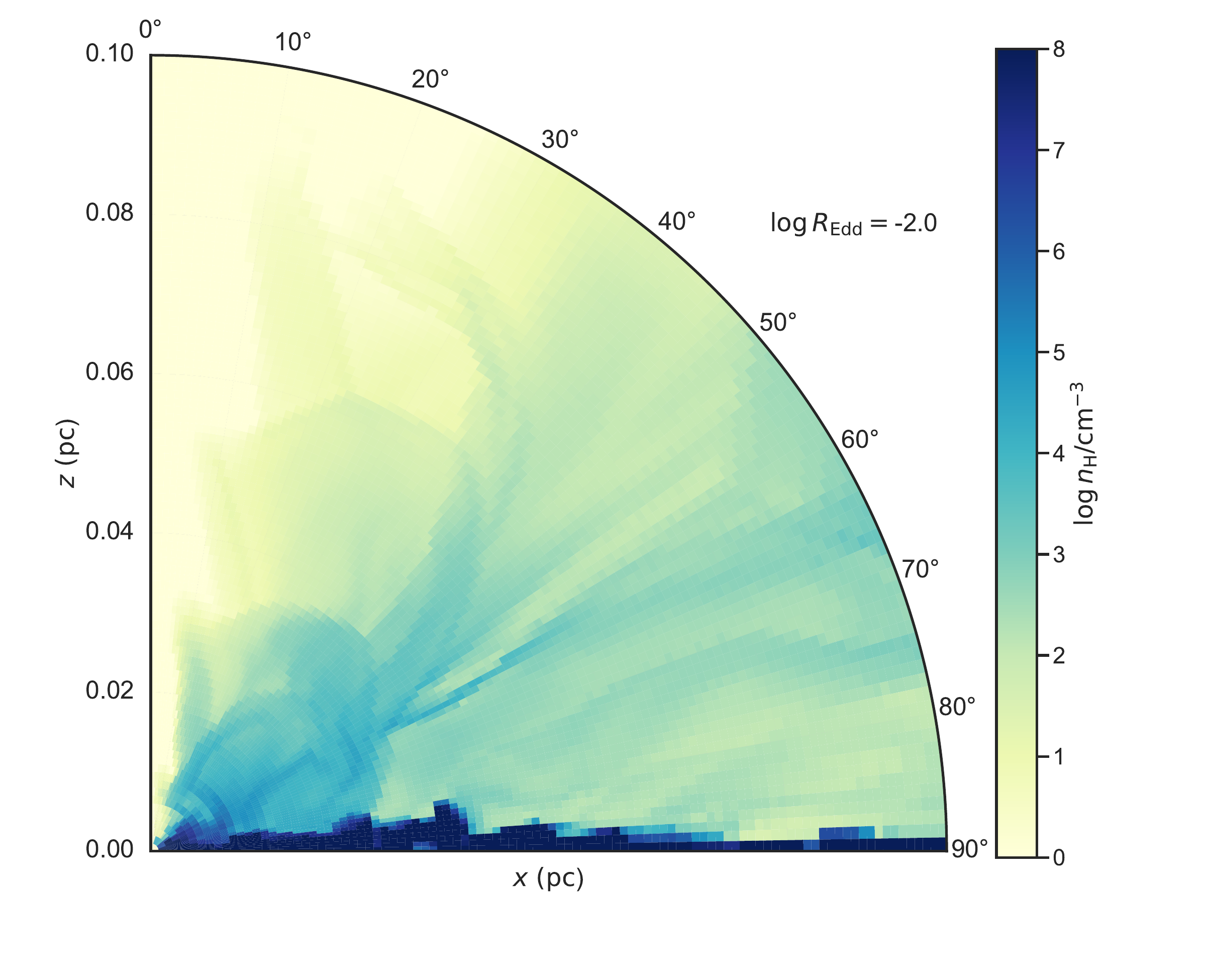}{0.5\textwidth}{(b)}}
\gridline{\fig{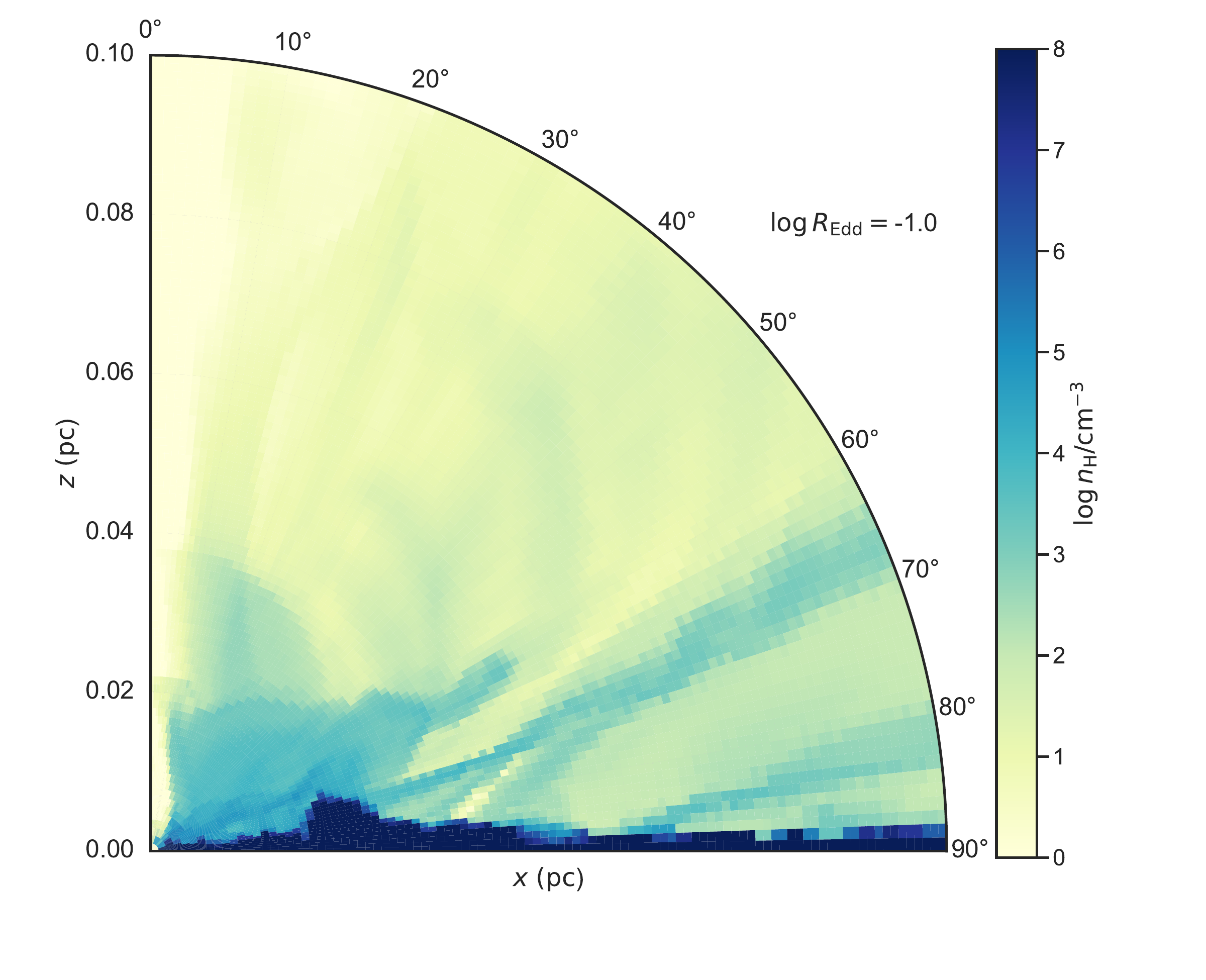}{0.5\textwidth}{(c)}\fig{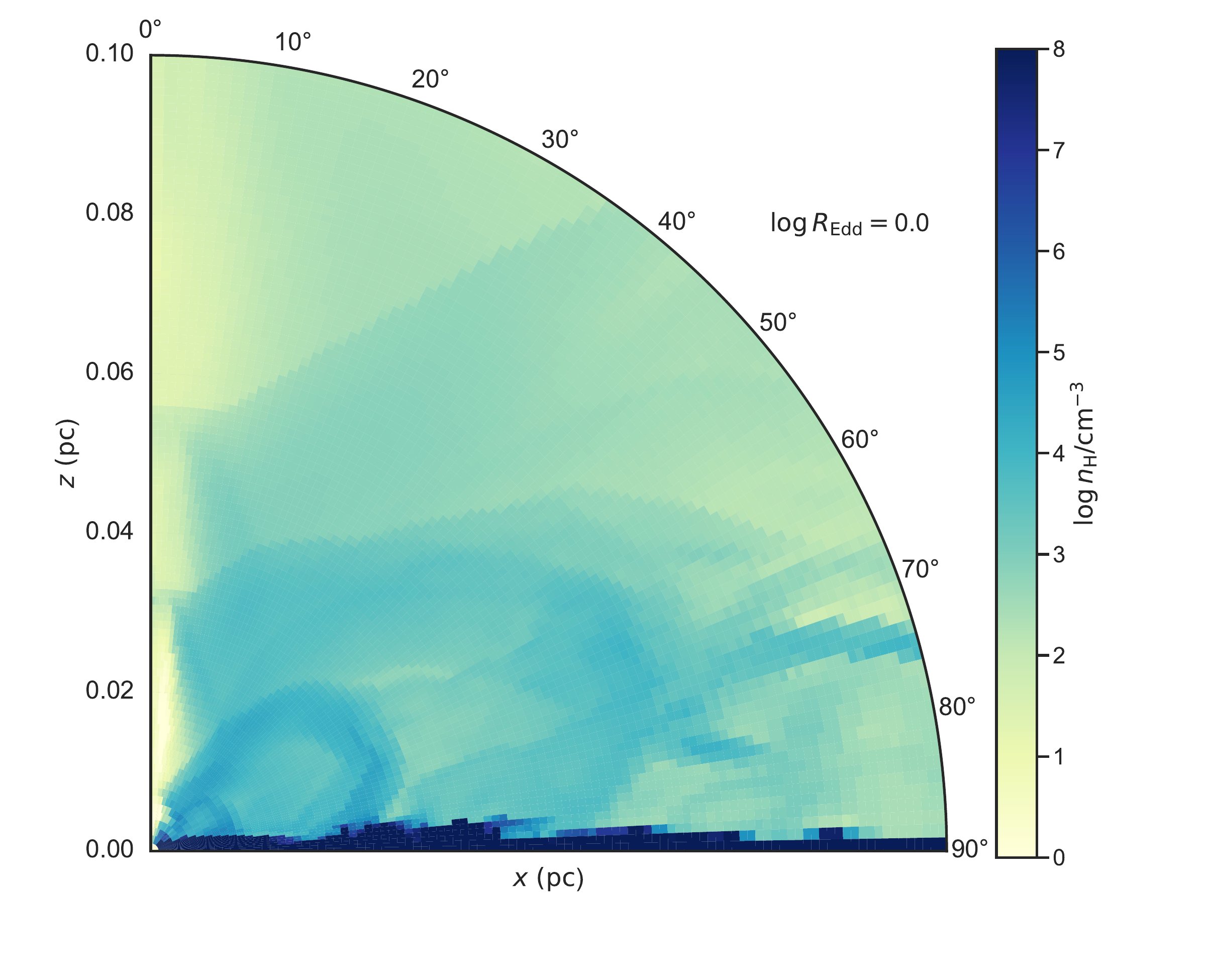}{0.5\textwidth}{(d)}}
\caption{(a) The distribution of the hydrogen number density $\log n_{\mathrm{H}}/\mathrm{cm}^{-3}$ obtained from the radiative hydrodynamic simulation with logarithmic Eddington ratio $\log R_{\mathrm{Edd}}$ of $-3$. (b) The $\log n_{\mathrm{H}}/\mathrm{cm}^{-3}$ distribution  with $\log R_{\mathrm{Edd}} = -2$. (c) The $\log n_{\mathrm{H}}/\mathrm{cm}^{-3}$ distribution with $\log R_{\mathrm{Edd}} = -1$. (d) The $\log n_{\mathrm{H}}/\mathrm{cm}^{-3}$ distribution with $\log R_{\mathrm{Edd}} = 0$.}
\end{figure*}
\begin{figure*}\label{Figure0003}
\gridline{\fig{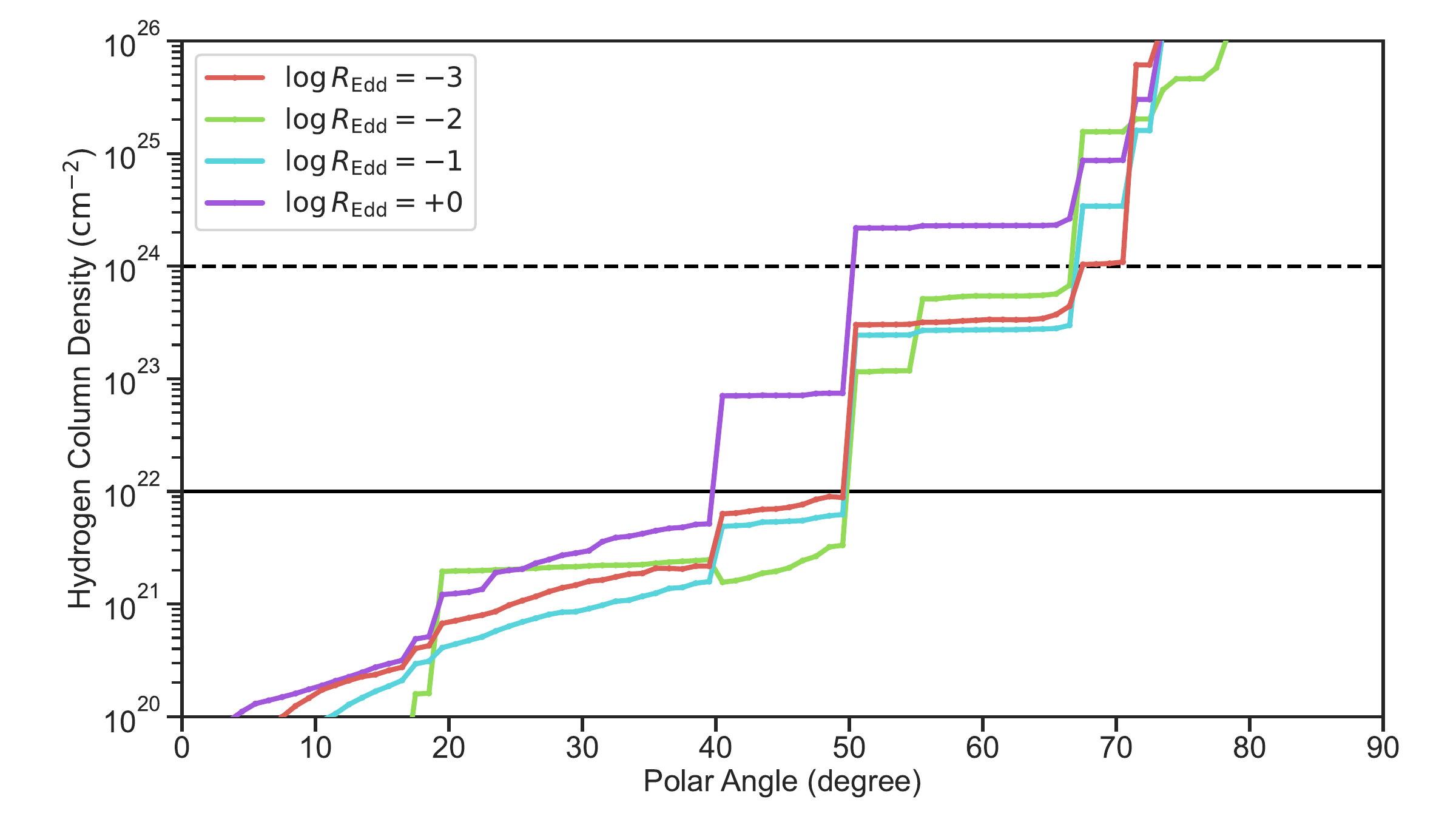}{0.5\textwidth}{}}
\caption{The averaged hydrogen column density as a function of the polar angle. The red, green, light blue, and purple lines correspond to the radiative hydrodynamic simulations of $\log R_{\mathrm{Edd}} = -3$, $-2$, $-1$, and $0$, respectively. The solid, and dashed lines represent the hydrogen column density of $10^{22} \ \mathrm{cm}^{-2}$ and $10^{24} \ \mathrm{cm}^{-2}$, respectively.}
\end{figure*}
\begin{figure*}\label{Figure0004}
\gridline{\fig{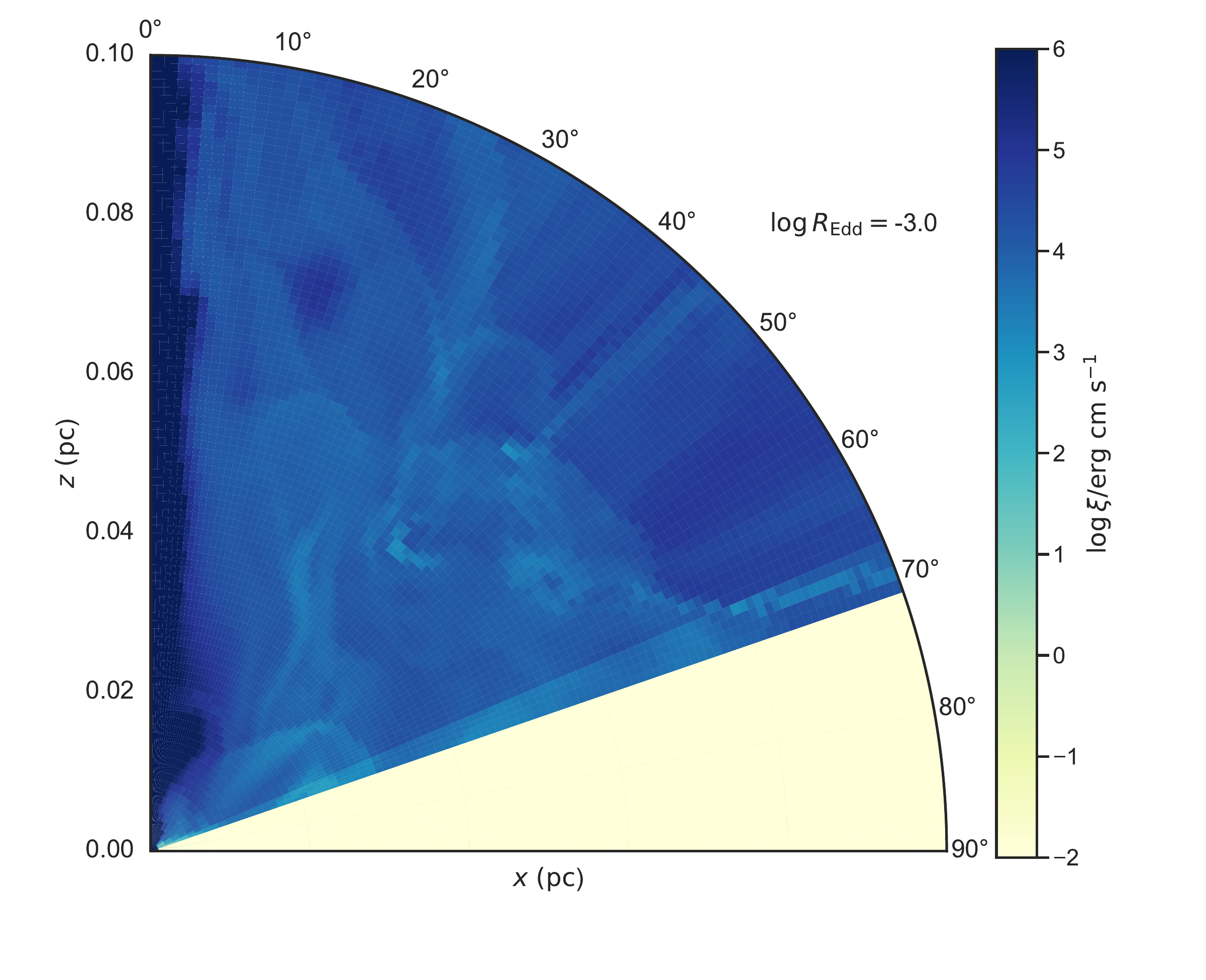}{0.5\textwidth}{(a)}\fig{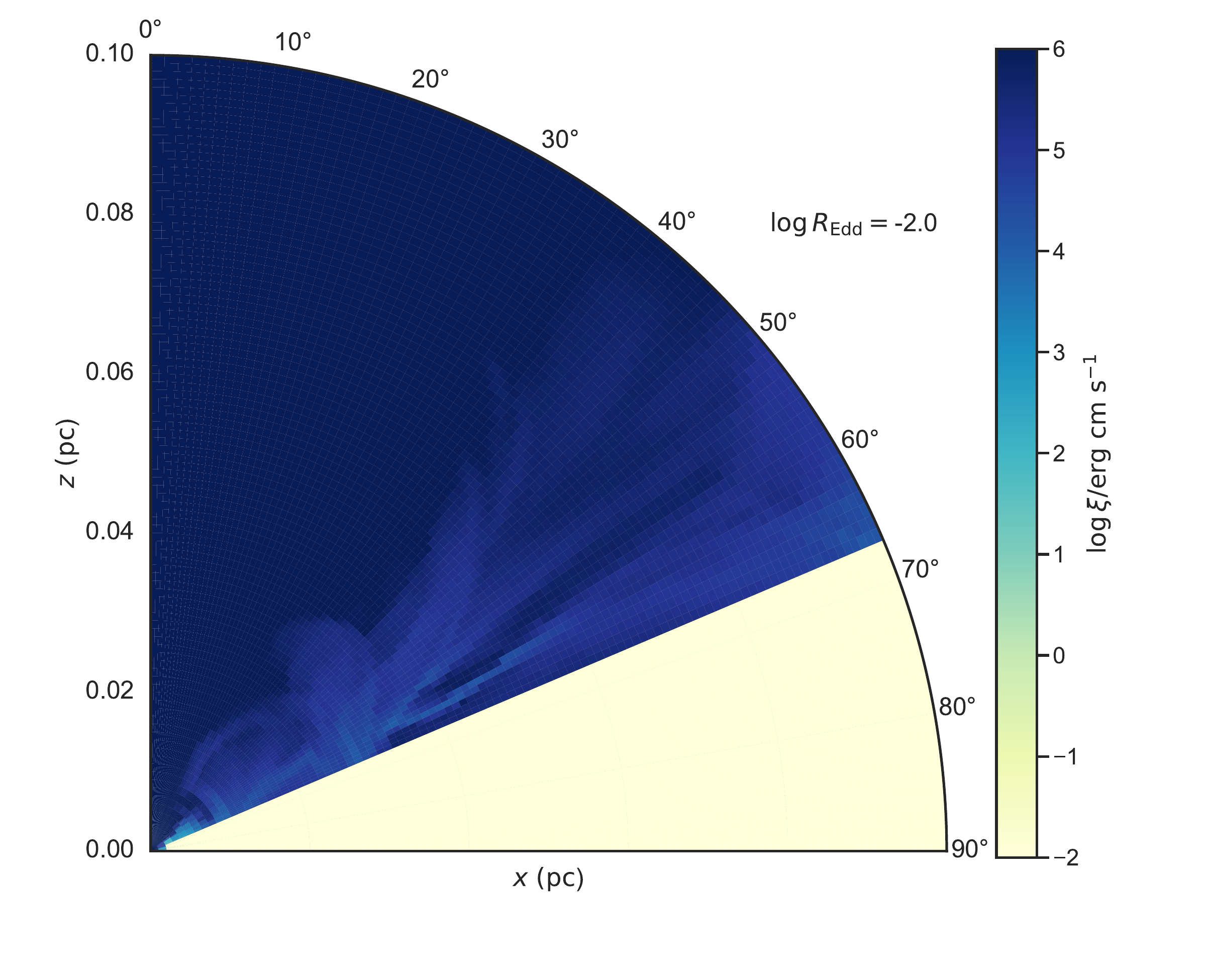}{0.5\textwidth}{(b)}}
\gridline{\fig{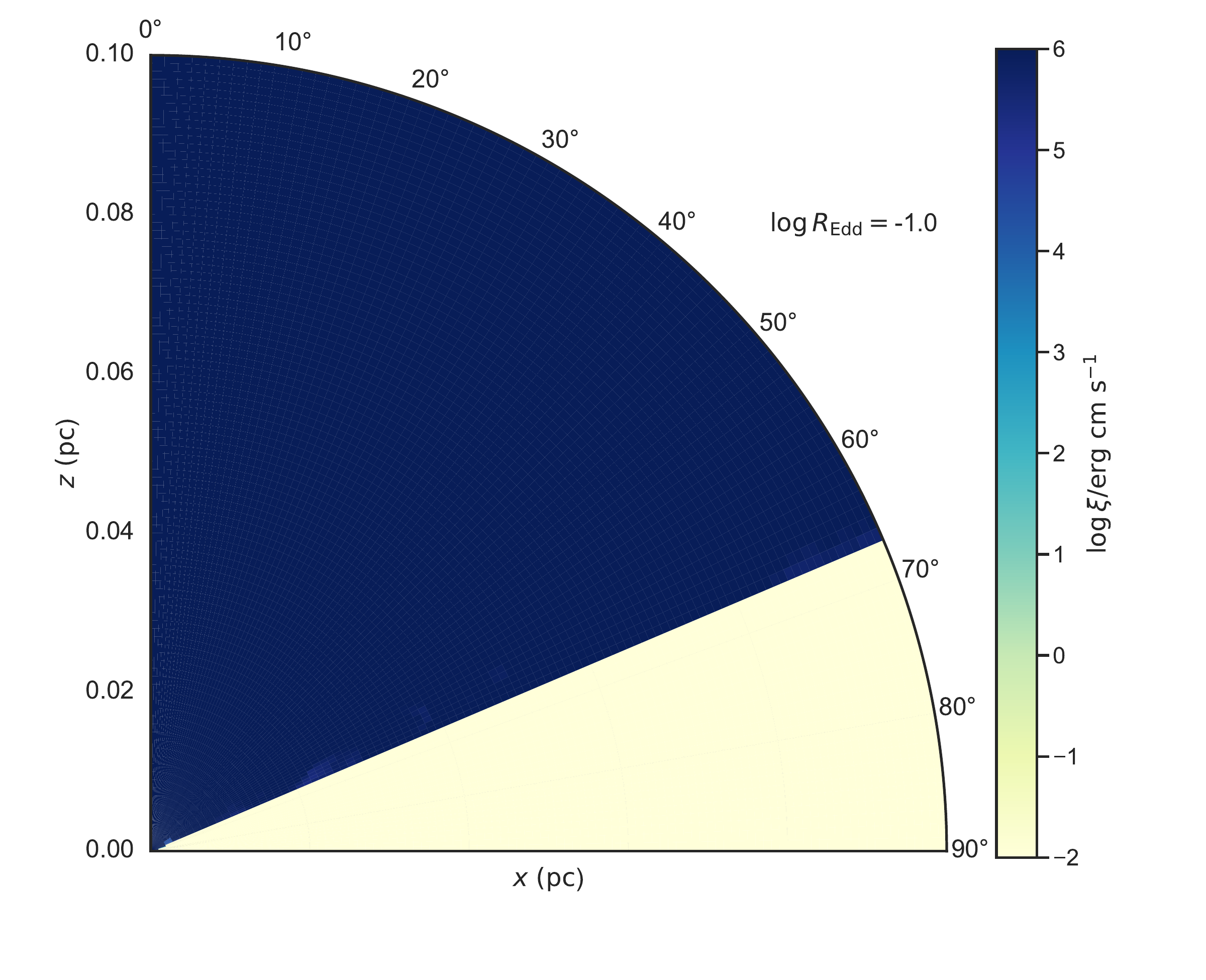}{0.5\textwidth}{(c)}\fig{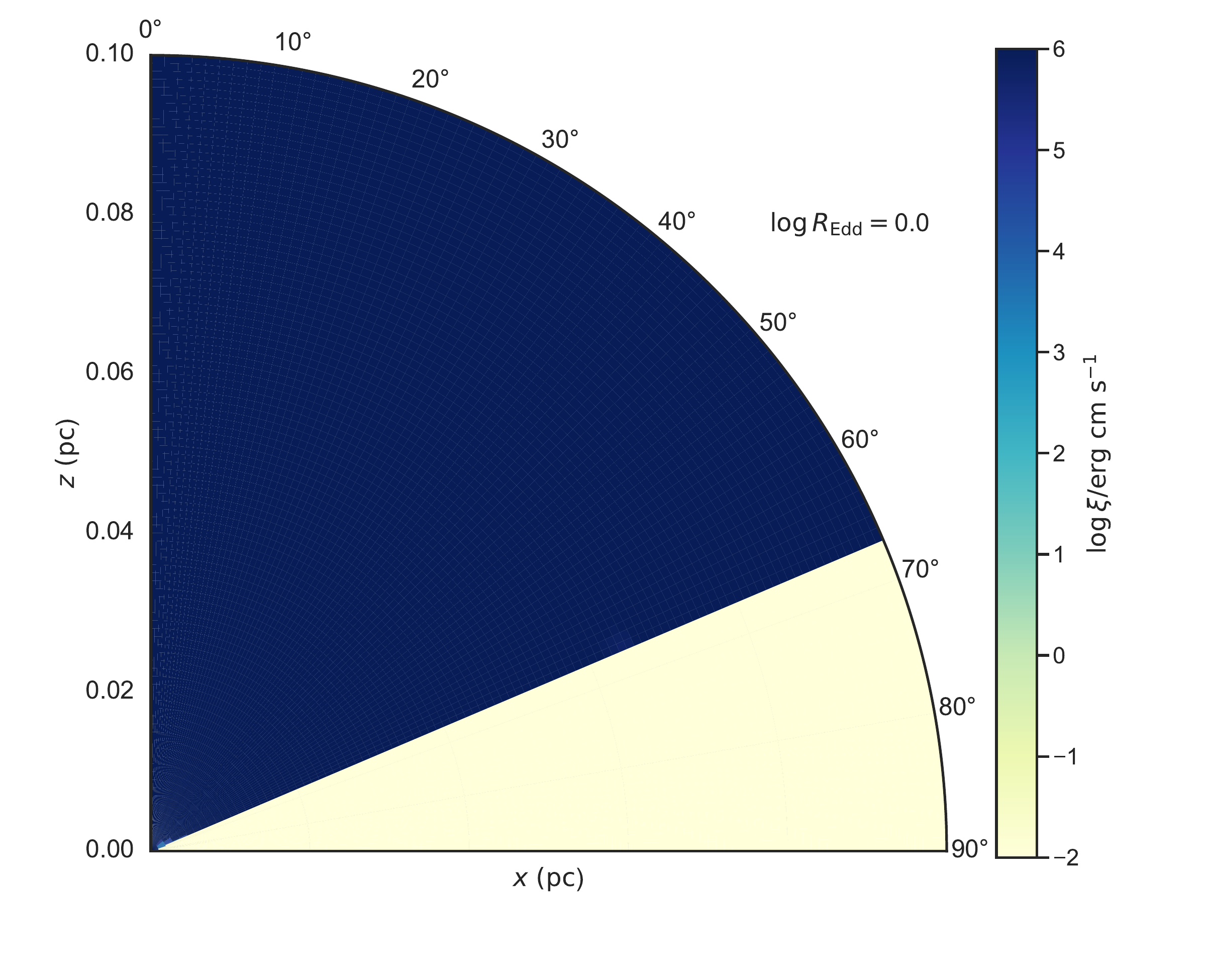}{0.5\textwidth}{(d)}}
\caption{(a) The distribution of the ionization parameter $\log \xi/\mathrm{erg} \ \mathrm{cm} \ \mathrm{s}^{-1}$ obtained from the radiative hydrodynamic simulation with logarithmic Eddington ratio $\log R_{\mathrm{Edd}}$ of $-3$. (b) The $\log \xi/\mathrm{erg} \ \mathrm{cm} \ \mathrm{s}^{-1}$ distribution with $\log R_{\mathrm{Edd}} = -2$. (c) The $\log \xi/\mathrm{erg} \ \mathrm{cm} \ \mathrm{s}^{-1}$ distribution with $\log R_{\mathrm{Edd}} = -1$. (d) The $\log \xi/\mathrm{erg} \ \mathrm{cm} \ \mathrm{s}^{-1}$ distribution with $\log R_{\mathrm{Edd}} = 0$.}
\end{figure*}
\begin{figure*}\label{Figure0005}
\gridline{\fig{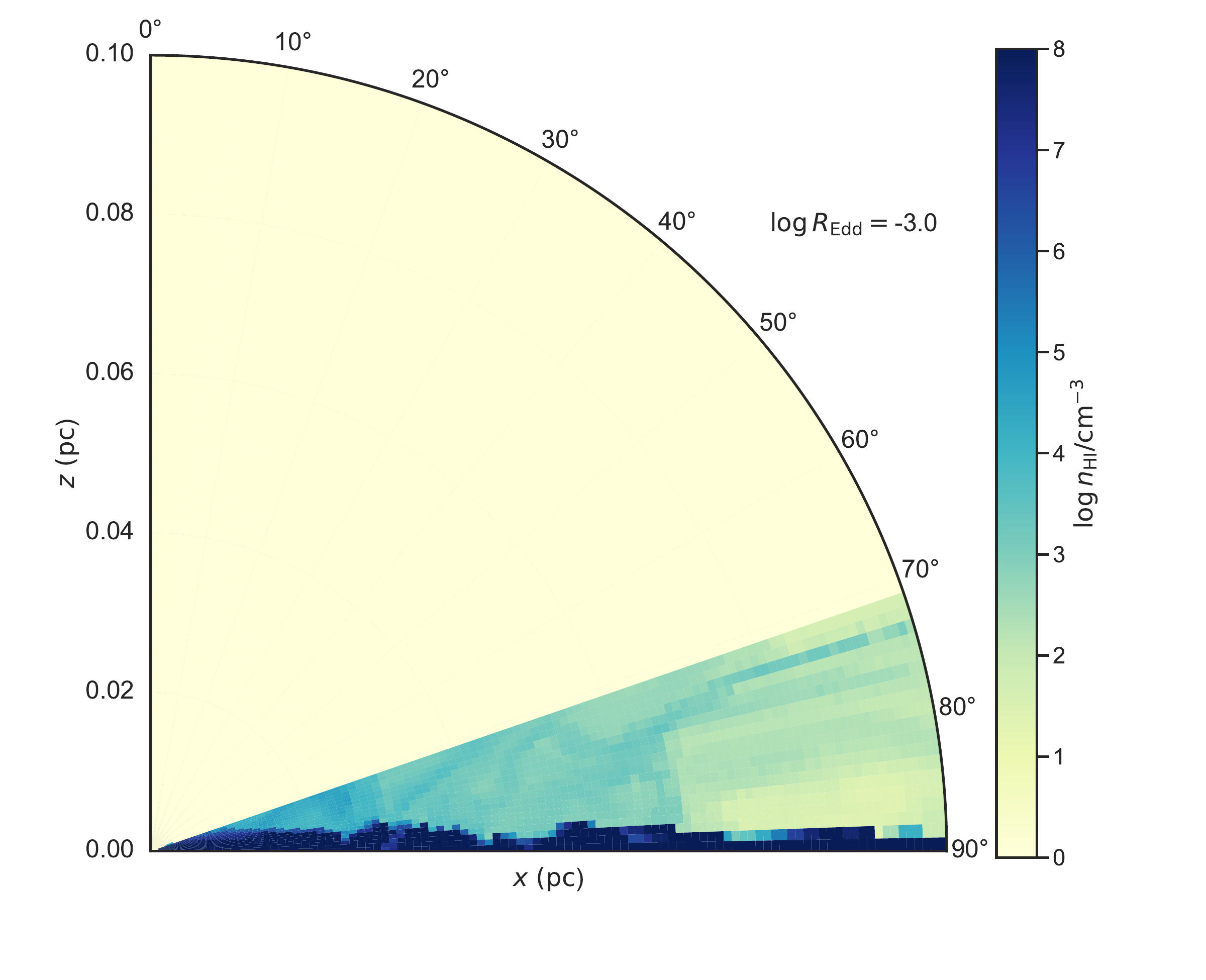}{0.5\textwidth}{(a)}\fig{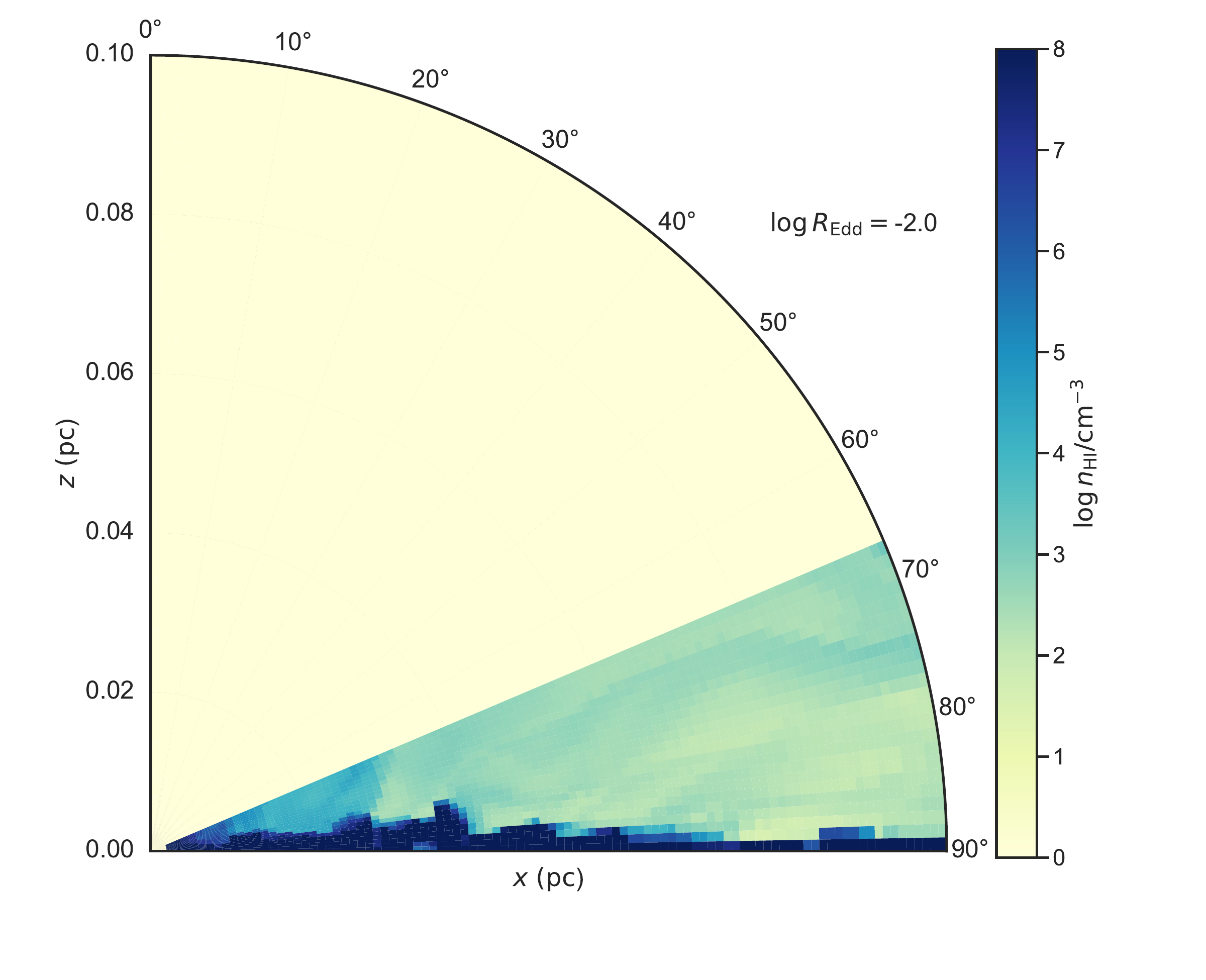}{0.5\textwidth}{(b)}}
\gridline{\fig{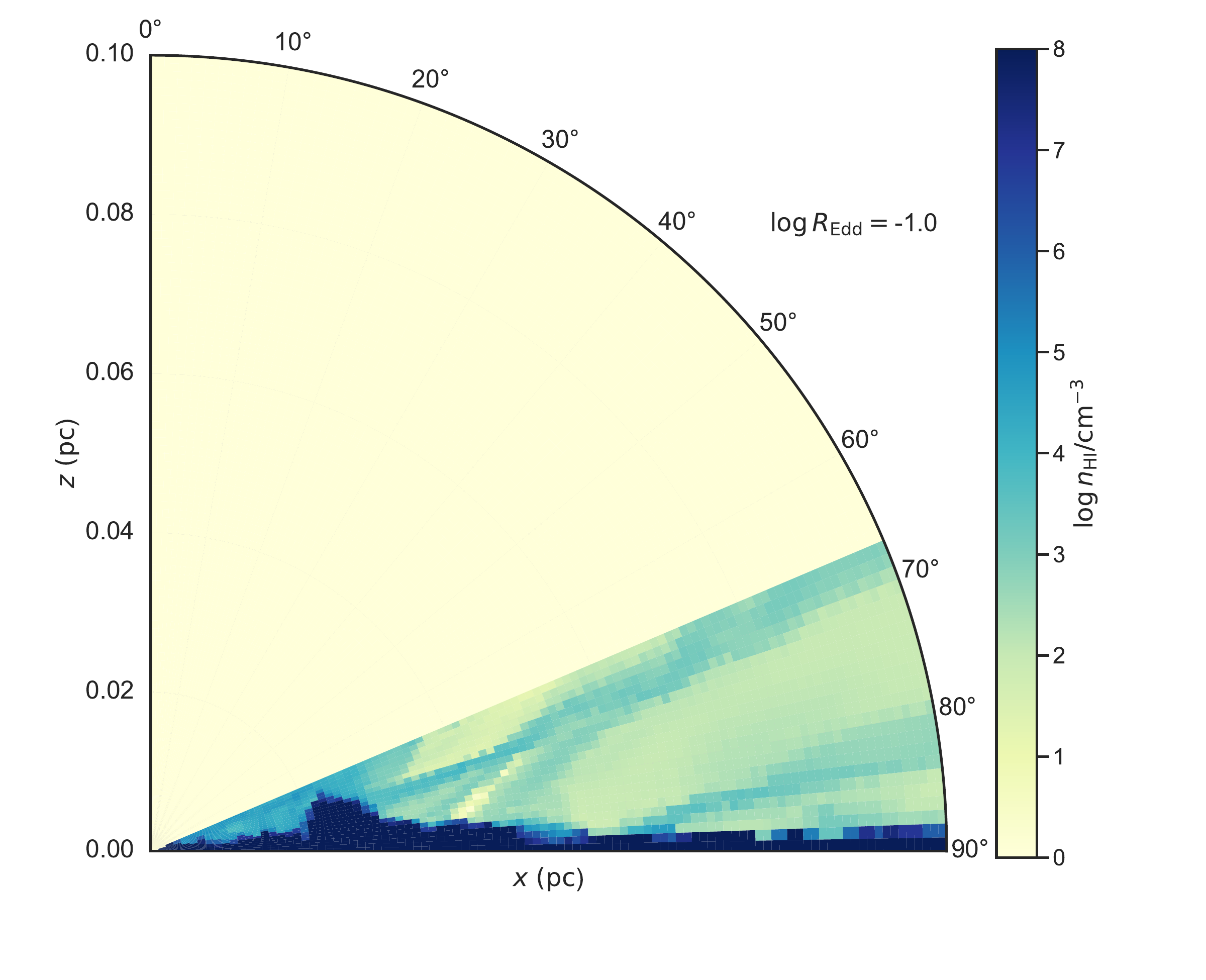}{0.5\textwidth}{(c)}\fig{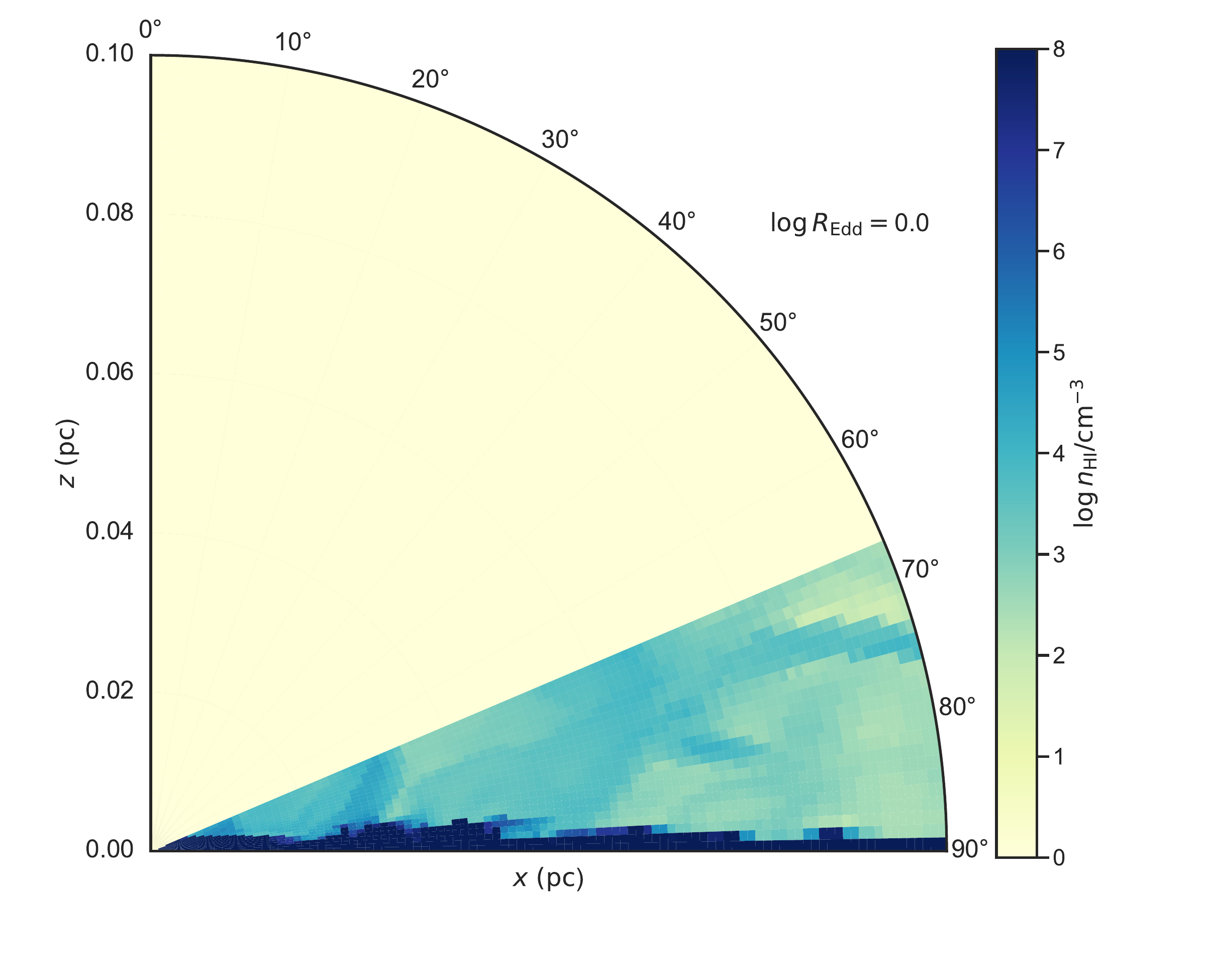}{0.5\textwidth}{(d)}}
\caption{(a) The distribution of \ion{H}{1} number density $\log n_{\mathrm{HI}}/\mathrm{cm}^{-3}$ obtained from the radiative hydrodynamic simulation with logarithmic Eddington ratio $\log R_{\mathrm{Edd}}$ of $-3$. (b) The $\log n_{\mathrm{HI}}/\mathrm{cm}^{-3}$ distribution with $\log R_{\mathrm{Edd}} = -2$. (c) The $\log n_{\mathrm{HI}}/\mathrm{cm}^{-3}$ distribution with $\log R_{\mathrm{Edd}} = -1$. (d) The $\log n_{\mathrm{HI}}/\mathrm{cm}^{-3}$ distribution with $\log R_{\mathrm{Edd}} = 0$.}
\end{figure*}
\begin{figure*}\label{Figure0006}
\gridline{\fig{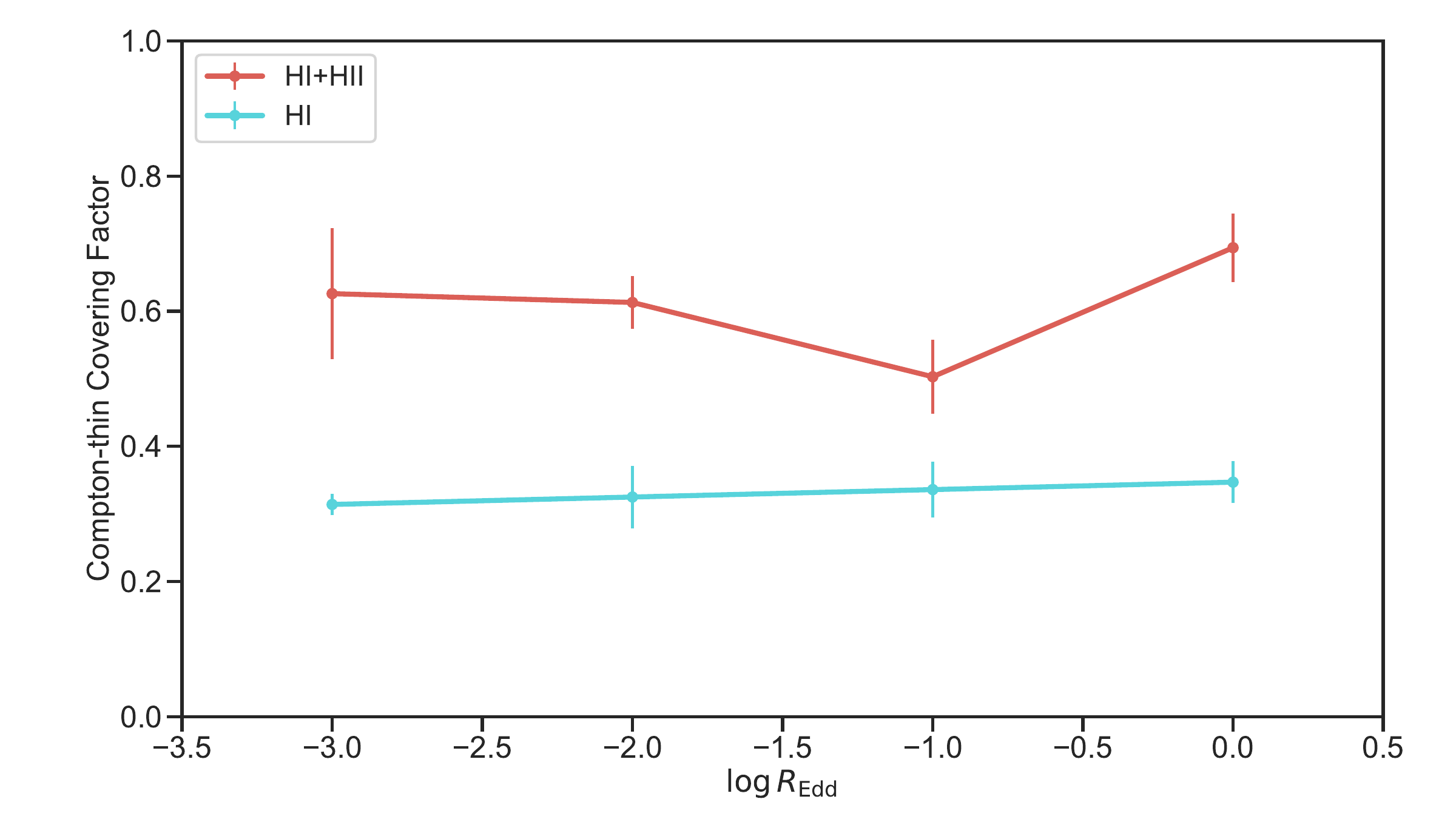}{0.5\textwidth}{(a)}\fig{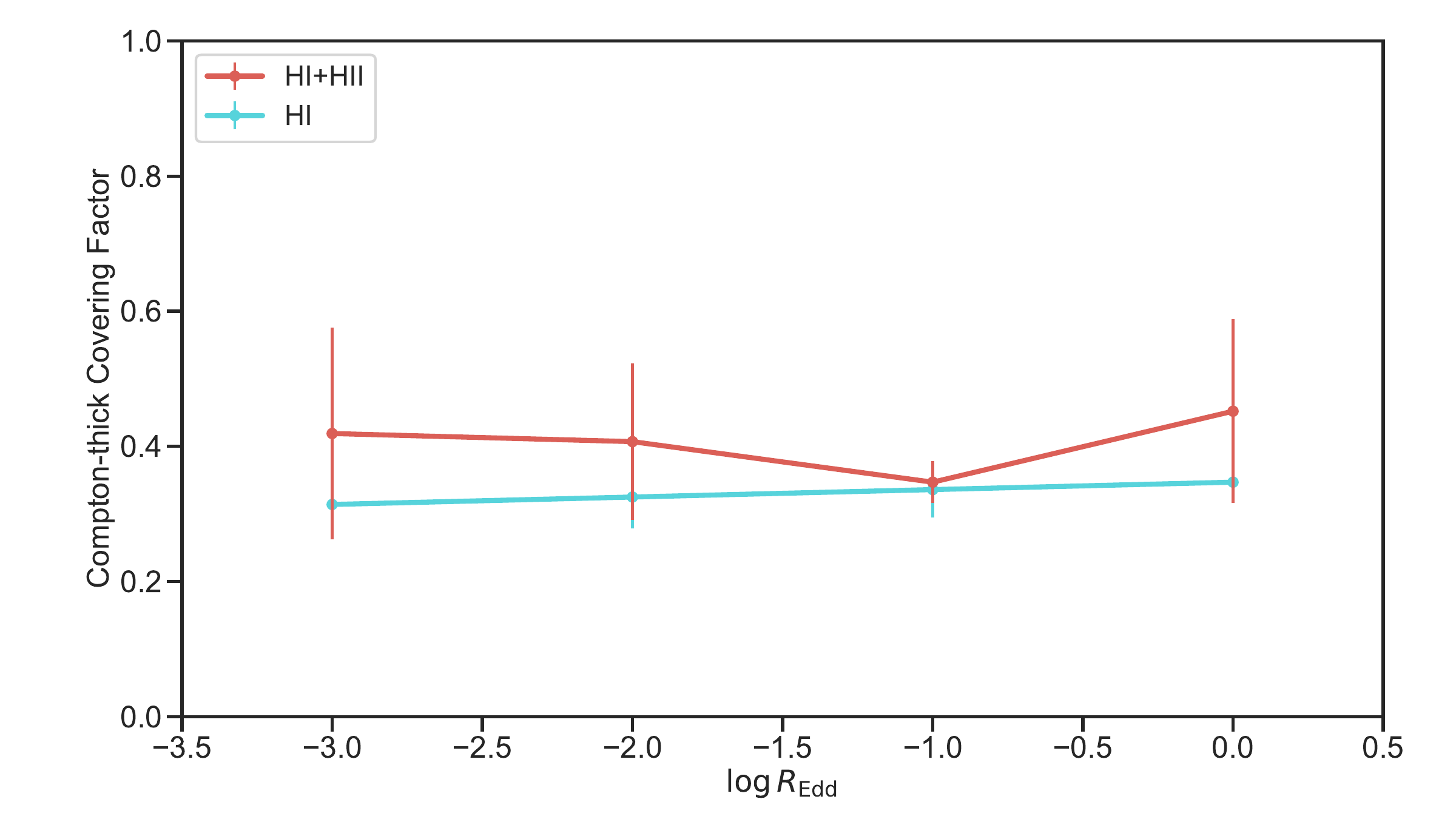}{0.5\textwidth}{(b)}}
\caption{(a) The relationship between the Eddington ratio and the Compton-thin covering factor $C_{22}$. The red line represents \ion{H}{1} and \ion{H}{2} $C_{22}$. The light blue line corresponds to \ion{H}{1} $C_{22}$. (b) The relationship between the Eddington ratio and the Compton-thick covering factor $C_{24}$. The red line represents \ion{H}{1} and \ion{H}{2} $C_{24}$. The light blue line corresponds to  \ion{H}{1} $C_{24}$.}
\end{figure*}\clearpage
\section{Results}\label{Section0300}
\hyperref[Section0301]{Section~3.1} presents the dependence of the hydrogen number density distribution on the Eddington ratio. \hyperref[Section0302]{Section~3.2} shows the dependence of the distribution of the ionization parameter on the Eddington ratio. \hyperref[Section0303]{Section~3.3} presents the dependence of the \ion{H}{1} number density distribution on the Eddington ratio. \hyperref[Section0304]{Section~3.4} shows the \ion{H}{1} Compton-thin covering factor and the \ion{H}{1} Compton-thick covering factor as a function of the Eddington ratio.
\subsection{Dependence of Hydrogen Number Density Distribution on Eddington Ratio}\label{Section0301}
We studied the dependence of the hydrogen number density $\log n_{\mathrm{H}}/\mathrm{cm}^{-3}$ distribution on the Eddington ratio. \hyperref[Figure0002]{Figure~2} show $\log n_{\mathrm{H}}/\mathrm{cm}^{-3}$ distributions obtained from radiative hydrodynamic simulations with (a) the logarithmic Eddington ratio $\log R_{\mathrm{Edd}} = -3$, (b) $\log R_{\mathrm{Edd}} = -2$, (c) $\log R_{\mathrm{Edd}} = -1$, and (d) $\log R_{\mathrm{Edd}} = 0$. All density distributions consist of dense material in the equatorial plane with $\log n_{\mathrm{H}}/\mathrm{cm}^{-3} \simeq 8$ and outflow with $\log n_{\mathrm{H}}/\mathrm{cm}^{-3} \simeq 4$.

The physical quantity obtained from the observation is not the hydrogen number density, but the hydrogen column density. The hydrogen number density distribution obtained from simulations evolves over time. To investigate the averaged hydrogen column density, we averaged the hydrogen number density distributions obtained from snapshots at $t = 0,~30,~60,~\mathrm{and}~90~\mathrm{years}$. \hyperref[Figure0003]{Figure~3} shows the averaged hydrogen column density as a function of the polar angle $\theta$. In the case of $\log R_{\mathrm{Edd}} = -3,~-2,~\mathrm{and}~-1$, the hydrogen column density reaches $10^{22} \ \mathrm{cm}^{-2}$ at $\theta = 50\degr$. In the case of $\log R_{\mathrm{Edd}} = 0$, the hydrogen column density reaches $10^{22} \ \mathrm{cm}^{-2}$ at $\theta = 40\degr$. That is, the relationship between the polar angle and the hydrogen column density does not change significantly even as the Eddington ratio increases. If the hydrogen column density along the line of sight exceeds $10^{25} \ \mathrm{cm}^{-2}$, even X-rays cannot penetrate it. Since all models have hydrogen column densities exceeding $10^{25} \ \mathrm{cm}^{-2}$ at polar angle between $65\degr$ and $75\degr$, regions with polar angle greater than these values are expected to be almost neutral.
\subsection{Dependence of Ionization Parameter Distribution on the Eddington Ratio}\label{Section0302}
We perform one-dimensional photoionization equilibrium calculations based on radiative hydrodynamic simulations. \hyperref[Figure0004]{Figure~4} shows the distributions of the ionization parameters $\log \xi/\mathrm{erg} \ \mathrm{cm} \ \mathrm{s}^{-1}$ with (a) $\log R_{\mathrm{Edd}} = -3$, (b) $\log R_{\mathrm{Edd}} = -2$, (c) $\log R_{\mathrm{Edd}} = -1$, and (d) $\log R_{\mathrm{Edd}} = 0$. Here, the ionization parameter $\xi$ is defined as $\xi = L/n_{\mathrm{H}}r^2$, where $L$ is the luminosity in $1$--$1000~\mathrm{Ryd}$, $n_{\mathrm{H}}$ is the hydrogen number density, and $r$ is the distance from the central source. 

\hyperref[Figure0004]{Figure~4} indicates that the distribution of $\log \xi/\mathrm{erg} \ \mathrm{cm} \ \mathrm{s}^{-1}$ varies greatly between $\theta$ of $65\degr$ and $75\degr$. If $\theta$ is less than $70\degr$, the ionization parameters are proportional to the luminosity of the central source because the line of sight is optically thin. In particular, for models with a low Eddington ratio, this region exhibits a moderate ionization parameter with $\log \xi/\mathrm{erg} \ \mathrm{cm} \ \mathrm{s}^{-1} = $2--4, which makes it likely to be observed as warm absorbers. Since we plan to perform three-dimensional X-ray radiative transfer calculations based on these models in future papers, we will not address warm absorbers in this paper. When $\theta$ is greater than $70\degr$, the ionization parameters are independent of the Eddington ratio and are nearly neutral. This is because, as explained in \hyperref[Section0301]{Section~3.1}, the line of sight is optically thick and the X-rays are almost completely absorbed.\clearpage
\subsection{Dependence of Neutral Hydrogen Number Density Distribution on the Eddington Ratio}\label{Section0303}
To calculate the covering factor for the neutral material, the distribution of \ion{H}{1} number density is required. Since we obtain \ion{H}{1} and \ion{H}{2} fractions for each region from XSTAR calculations, we define the \ion{H}{1} number density in each region as the hydrogen number density multiplied by the \ion{H}{1} fraction. \hyperref[Figure0005]{Figure~5} shows the distribution of the \ion{H}{1} number density $\log n_{\mathrm{HI}}/\mathrm{cm}^{-3}$ with (a) $\log R_{\mathrm{Edd}} = -3$, (b) $\log R_{\mathrm{Edd}} = -2$, (c) $\log R_{\mathrm{Edd}} = -1$, and (d) $\log R_{\mathrm{Edd}} = 0$. \hyperref[Figure0005]{Figure~5} shows only the region in \hyperref[Figure0002]{Figure~2} where the logarithmic ionization parameter is less than $-2$. The reason is that, for \ion{H}{1} to exist, the logarithmic ionization parameter must be less than $-2$ \citep[][Figure~1]{Kallman1982}. As explained in \hyperref[Section0302]{Section~3.2}, the region where $\theta$ is greater than $70\degr$ is nearly neutral (\hyperref[Figure0004]{Figure~4}). \hyperref[Figure0005]{Figure~5} indicates that \ion{H}{1} can survive only in the region where $\theta$ is greater than $70\degr$.

We note that even if all \ion{H}{1} becomes \ion{H}{2}, it does not mean that the material around the AGN will not absorb X-rays through photoelectric absorption. This is because the material around AGNs not only contains hydrogen but also helium, carbon, oxygen, neon, magnesium, silicon, sulfur, iron, nickel, and others. When estimating the hydrogen column density along the line of sight from X-ray spectroscopic observations, the X-ray spectrum in the $2$--$8~\mathrm{keV}$ energy band is frequently utilized. In this energy band, photoelectric absorption by \ion{O}{1} is the primary process \citep{Morrison1983, Vander2023}. When the logarithmic ionization parameter is greater than $2$, since almost all oxygen becomes \ion{O}{9}, X-ray absorption due to photoelectric absorption becomes very weak \citep[][Figure~1]{Kallman1982}. This does not affect the results of this paper. The reason is that the logarithmic ionization parameter has discrete values of $-2$ and $+2$ around $\theta = 70\degr$ (\hyperref[Figure0004]{Figure~4}).
\subsection{Covering Factor of Neutral Hydrogen}\label{Section0304}
We calculated the \ion{H}{1} covering factor based on radiative hydrodynamic simulations and photoionization equilibrium calculations. \hyperref[Figure0006]{Figure~6(a)} compares $C_{22}$ of \ion{H}{1} alone with $C_{22}$ of \ion{H}{1} and \ion{H}{2}. Since the density structures in the simulations vary over time, we obtain the covering factor averaged in time from four snapshots at $t = 0, \ 30, \ 60, \mathrm{and} \ 90 \ \mathrm{years}$. Here, the error corresponds to $1\sigma$. The Compton-thin covering factors $C_{22}$ of \ion{H}{1} and \ion{H}{2} are almost independent of the Eddington ratio and are approximately $70\%$. The Compton-thin covering factors $C_{22}$ of \ion{H}{1} alone are also independent of the Eddington ratio and are approximately $30\%$. Since the region within $0.1~\mathrm{pc}$ is photoionized even in an Eddington ratio of $10^{-3}$, $C_{22}$ of \ion{H}{1} is independent of the Eddington ratio. \hyperref[Figure0006]{Figure~6(b)} compares $C_{24}$ of \ion{H}{1} alone with $C_{24}$ of \ion{H}{1} and \ion{H}{2}. The Compton-thick covering factors $C_{24}$ of \ion{H}{1} and \ion{H}{2} are smaller than $C_{22}$ and are approximately $40\%$. The Compton-thick covering factors $C_{24}$ of \ion{H}{1} alone are independent of the Eddington ratio and are approximately $30\%$. In the case of the region under consideration, since all Compton-thin gasses are photoionized, \ion{H}{1} $C_{24}$ has the same value as \ion{H}{1} $C_{22}$.\clearpage
\begin{figure*}\label{Figure0007}
\gridline{\fig{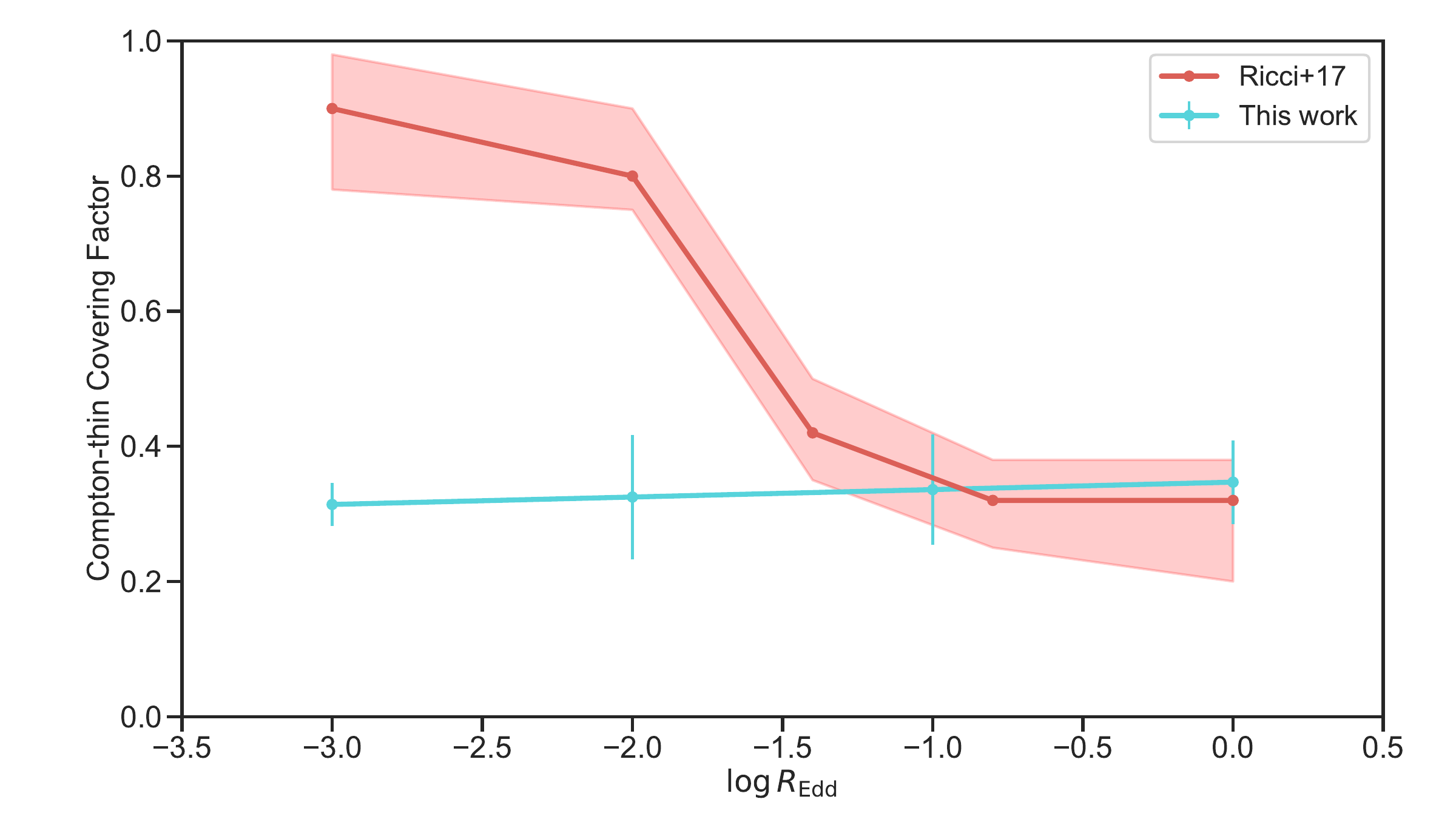}{0.5\textwidth}{(a)}\fig{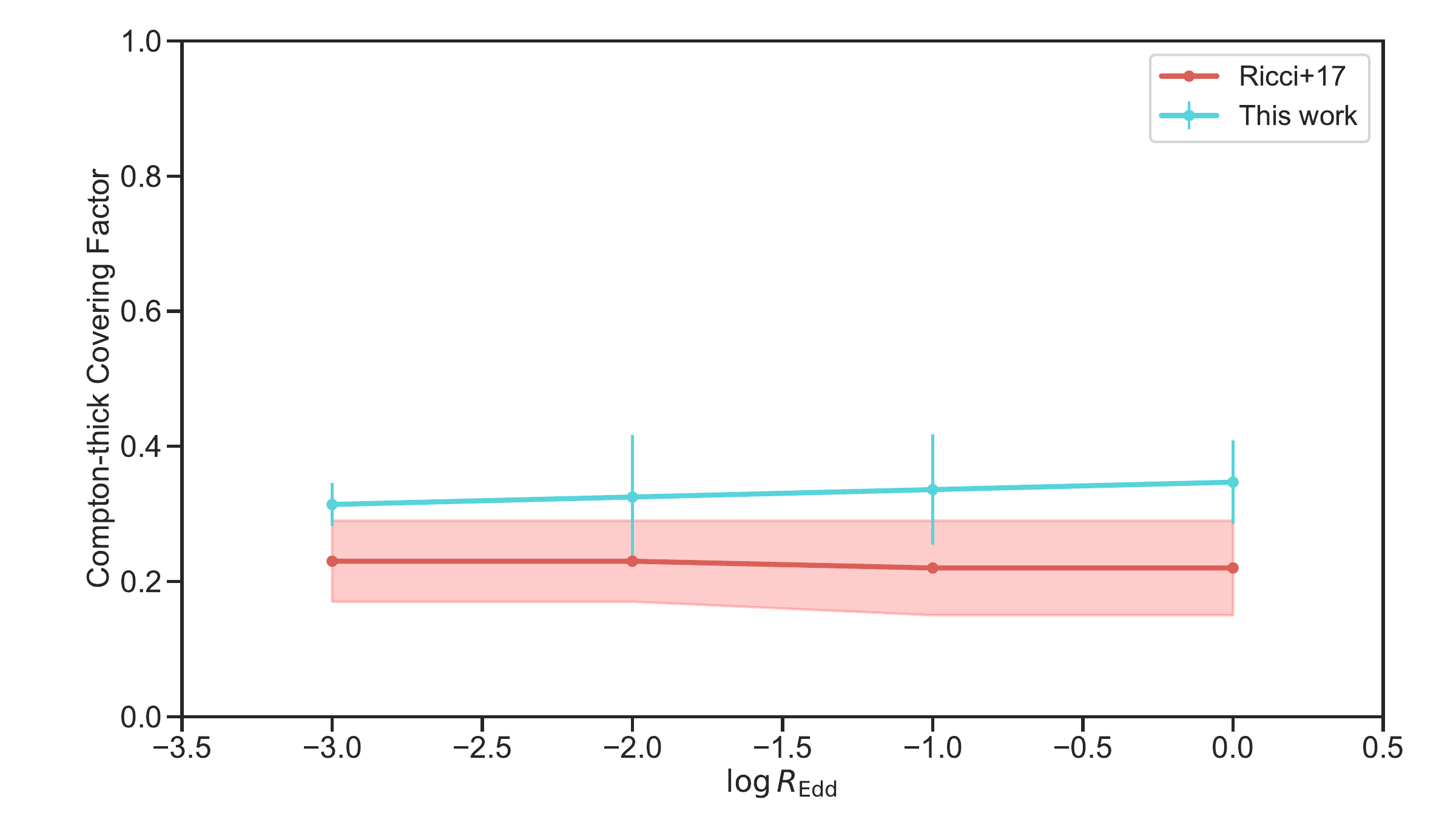}{0.5\textwidth}{(b)}}
\caption{(a) The comparison between the Compton-thin covering factor $C_{22}$ obtained from our simulation and that inferred from X-ray observations. The red line represents $C_{22}$ obtained from the X-ray observations \citep{Ricci2017a}. The light blue line corresponds to $C_{22}$ obtained from our simulation. (b) The comparison between the Compton-thick covering factor $C_{24}$ obtained from our simulation and that inferred from the X-ray observations. The red line represents $C_{24}$ obtained from the X-ray observations. The light blue line corresponds to $C_{24}$ obtained from our simulation.}
\end{figure*}
\begin{figure*}\label{Figure0008}
\gridline{\fig{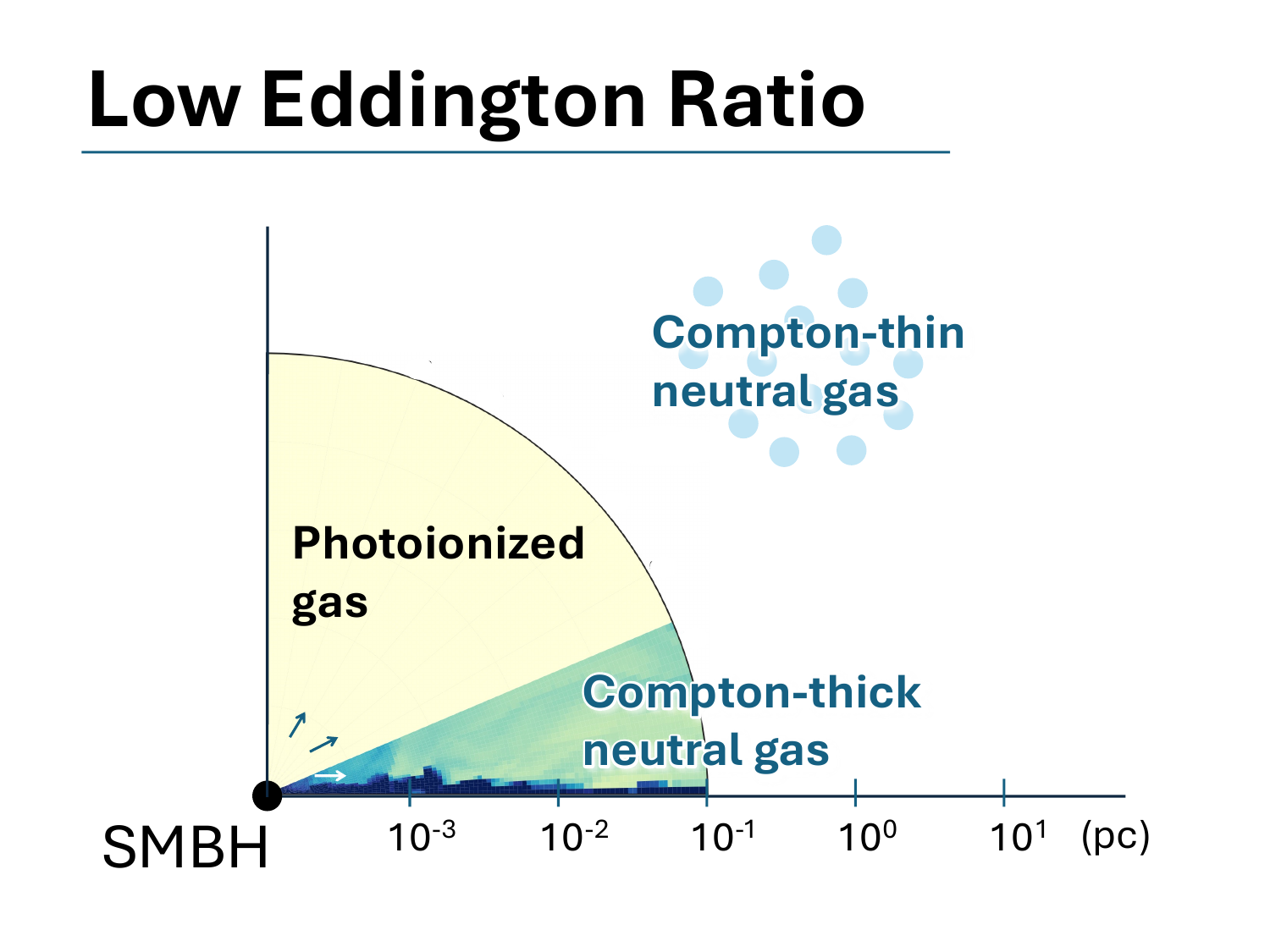}{0.5\textwidth}{(a)}\fig{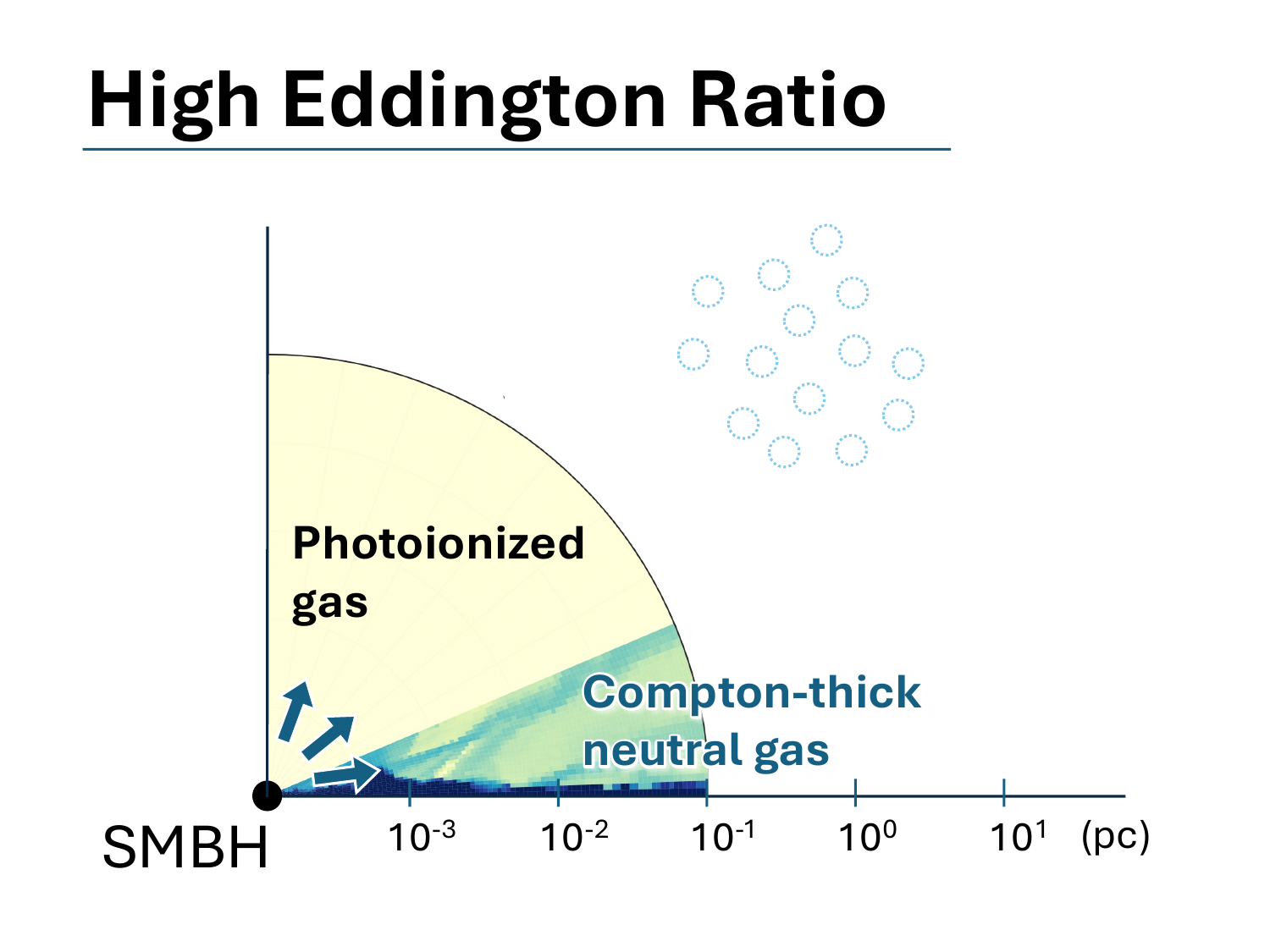}{0.5\textwidth}{(b)}}
\caption{Schematic diagram of a unified model of AGN based on photoionization equilibrium. In both cases, Compton-thick neutral gas with a covering factor of approximately $30\%$ exists at a distance of approximately $10^{-3}$--$10^{-1}~\mathrm{pc}$. If the Eddington ratio is less than $10^{-2}$, Compton-thin neutral gas exists at a distance of approximately $10~\mathrm{pc}$. When the Eddington ratio is greater than $10^{-1}$, since the Compton-thin neutral gas is photoionized, the covering factor of the Compton-thin gas decreases.}
\end{figure*}\clearpage
\section{Discussion}\label{Section0400}
To estimate the covering factor of \ion{H}{1}, we performed photoionization equilibrium calculations based on radiative hydrodynamic simulations. \hyperref[Section0401]{Section~4.1} compares the covering factor obtained from our simulation with that inferred from the X-ray observations. \hyperref[Section0402]{Section~4.2} discusses the physical mechanisms that determine the covering factor. \hyperref[Section0403]{Section~4.3} proposes an AGN unified model based on photoionization.
\subsection{Comparison of Covering Factors}\label{Section0401}
We compare the covering factor obtained from theoretical calculations with that inferred from X-ray observations. \hyperref[Figure0007]{Figure~7(b)} compares the results of $C_{24}$ obtained from our simulations with that inferred from the X-ray observations \citep{Ricci2017a}. As explained in \hyperref[Section0304]{Section~3.4}, our $C_{24}$ is independent of the Eddington ratio and is approximately $30\%$. Our $C_{24}$ is consistent with that obtained from the X-ray observations \citep{Ricci2017a, Tanimoto2022, Boorman2025}. In other words, our model, which considers the region from $10^{-3}$ to $10^{-1}~\mathrm{pc}$, can explain $C_{24}$ obtained from X-ray observations.

\hyperref[Figure0007]{Figure~7(a)} compares the results of $C_{22}$ obtained from our simulations with that inferred from the X-ray observations. Our $C_{22}$ is independent of the Eddington ratio and is approximately $30\%$. When the Eddington ratio is greater than $10^{-1}$, our $C_{22}$ is consistent with that obtained from the X-ray observations. If the Eddington ratio is smaller than $10^{-2}$, our $C_{22}$ is smaller than that obtained from the X-ray observations. This is because even if Compton-thin gas exists inside $0.1~\mathrm{pc}$, it is photoionized by X-ray radiation from the central source.
\subsection{Physical Mechanism Determining Covering Factor}\label{Section0402}
The radiation-regulated AGN unification model considers that the radiation pressure blows off a Compton-thin dusty gas when the Eddington ratio is greater than $10^{-2}$. Although AGNs radiation plays a significant role as the physical mechanism that determines the covering factor, it remains unclear whether the primary effect is the radiation pressure blowing away matter or the photoionization of matter by radiation.

To distinguish them, we investigated the dependence of the hydrogen column density on the polar angle. We can directly examine the relationship between the polar angle and the hydrogen column density based on radiative hydrodynamic simulations. As shown in \hyperref[Figure0003]{Figure~3}, the relationship between the polar angle and the hydrogen column density does not change significantly even when the Eddington ratio changes inside $0.1~\mathrm{pc}$. Our result is consistent with \cite{Kudoh2024}. \cite{Kudoh2024} showed the hydrogen column densities as a function of the elevation angle for dusty and dust-free gasses. They suggested that the hydrogen column density of dusty gas decreases with the Eddington ratio, whereas that of dust-free gas is nearly independent of the Eddington ratio.

In their calculations, the hydrogen column density is primarily determined by hot, dust-free ionized gas located inside the dust sublimation radius. Since dusty gas located outside the sublimation radius is efficiently accelerated and expanded due to radiation pressure on the dust, their hydrogen column densities are not large. In contrast, the dust-free gas experiences weaker radiative acceleration and tends to remain gravitationally bound, maintaining relatively high number density \citep[][Figure~6]{Kudoh2023}. Hence, the hydrogen column density is dominated by dust-free gas rather than dust gas and is almost independent of the Eddington ratio \citep[][Figure~8]{Kudoh2024}. This is consistent with the observations of the nearby AGN \citep{Burtscher2016, Mizukoshi2024}. \cite{Mizukoshi2024} indicated that the hydrogen column density of dust-free gas is basically greater than that of dusty gas \citep[][Figure~4]{Mizukoshi2024}. That is, within the region less than $0.1~\mathrm{pc}$, the effect of photoionization is more significant than the effect of blowing away matter through radiation pressure.\clearpage
\subsection{AGN Unified Model via Photoionization}\label{Section0403}
Since our $C_{22}$ has the same value as our $C_{24}$, all AGNs are observed to be unobscured or Compton-thick AGNs. However, X-ray observations reveal that many Compton-thin AGNs are present \citep[e.g.,][]{Kawamuro2016}. To explain the existence of these Compton-thin AGNs, \ion{H}{1} must be present in regions beyond the current computational domain. If the Eddington ratio is $10^{-2}$, we can estimate the lower limit of the distance to the absorber under the condition that the Compton-thin \ion{H}{1} survives
\begin{align}
\begin{cases}
10^{22}~\mathrm{cm}^{-2} \leq N_{\mathrm{H}} & < 10^{24}~\mathrm{cm}^{-2}\\
\frac{L(R_{\mathrm{Edd}} = 10^{-2})}{n_{\mathrm{H}}r^2} & \leq 10^{-2}~\mathrm{erg}~\mathrm{cm}~\mathrm{s}^{-1}
\end{cases}
\end{align}
Here, $N_{\mathrm{H}}$ is the hydrogen column density along the line of sight, $L(R_{\mathrm{Edd}} = 10^{-2})$ is the X-ray luminosity at the Eddington ratio of $10^{-2}$, $n_{\mathrm{H}}$ is the hydrogen number density, and $r$ is the distance to the absorber. We assume that the hydrogen column density is expressed as $N_{\mathrm{H}} = n_{\mathrm{H}} r f$, where $f$ is the filling factor. Substituting the first equation of Equation~4 into the second equation of Equation~4 and simplifying produces the following lower limit for the distance to the absorber.
\begin{equation}
r \geq 30 \left(\frac{N_{\mathrm{H}}}{10^{22}~\mathrm{cm}^{-2}}\right)^{-1} \left(\frac{L}{10^{42} \ \mathrm{erg} \ \mathrm{s}^{-1}}\right) \ \left(\frac{f}{10^{-2}}\right) \ \mathrm{pc}.
\end{equation}
If the hydrogen column density is $10^{22}~\mathrm{cm}^{-2}$, the X-ray luminosity is $10^{42}~\mathrm{erg}~\mathrm{s}^{-1}$, and the filling factor is $10^{-2}$, the Compton-thin gas must exist beyond $30~\mathrm{pc}$. This radius extends beyond the computational domain of the current radiative hydrodynamic simulations. If a Compton-thin gas exists at this radius, it could explain the discrepancy between $C_{22}$ obtained from theoretical calculations and those derived from X-ray observations.

\hyperref[Figure0008]{Figure~8} shows a schematic diagram of the unified model of AGN based on photoionization equilibrium. A Compton-thick neutral gas with a covering factor of approximately $30\%$ exists at a distance of approximately $10^{-3}$--$10^{-1}~\mathrm{pc}$. If the Eddington ratio is smaller than $10^{-2}$, a Compton-thin neutral gas exists at a distance of approximately $30~\mathrm{pc}$. When the Eddington ratio is greater than $10^{-1}$, since the Compton-thin gas is photoionized, the covering factor becomes smaller. Note that the physical mechanism to blow off the Compton-thin gas to a large-scale height is an open question.
\section{Conclusion}\label{Section0500}
To clarify the relationship between the Eddington ratio and the covering factor, we performed the photoionization equilibrium calculations based on the two-dimensional axisymmetric radiative hydrodynamic simulations for four Eddington ratios. We calculated the Compton-thin covering factor $C_{22}$ and Compton-thick covering factor $C_{24}$ and compared those obtained from theoretical calculations with those inferred from X-ray observations. The key findings are as follows.

\begin{enumerate}
\item \ion{H}{1} and \ion{H}{2} $C_{22}$ is independent of the Eddington ratio and is approximately $70\%$. \ion{H}{1} $C_{22}$ is also independent of the Eddington ratio and is approximately $30\%$. This is because the Compton-thin gas within $0.1~\mathrm{pc}$ is photoionized by the radiation from the central source even at an Eddington ratio of $10^{-3}$.
\item We compared $C_{24}$ obtained from our simulation with that inferred from X-ray observations. Our $C_{24}$ is consistent with that inferred from the X-ray observations. This result suggests that the Compton-thick gases are formed by a radiation-driven fountain model.
\item Our $C_{22}$ agrees with that inferred from X-ray observations in a high Eddington ratio, whereas our $C_{22}$ is smaller than that from X-ray observations in a low Eddington ratio. To account for the large $C_{22}$ obtained from X-ray observations, a Compton-thin gas with a radius of approximately $10~\mathrm{pc}$ is required.
\item The relationship between the hydrogen column density and the polar angle is found to be almost independent of the Eddington ratio. This suggests that the covering factor is mainly determined by photoionization within the region inside $0.1~\mathrm{pc}$.
\end{enumerate}
\begin{acknowledgements}
The authors thank the anonymous referee who provided useful and detailed comments. Atsushi Tanimoto and the present research are supported by the Kagoshima University postdoctoral research program (KU-DREAM). This work is also supported by the Grant-in-Aid for Early Career Scientists Grant No. 23K13147 (A.T.), the Grants-in-Aid for Scientific Research Grant Nos. 25H00671 (K.W.), 24K17080 (Y.K.), and 22H00128 (H.O.), and the Exploratory Research Grant for Young Scientists, Hirosaki University (M.N.). Numerical computations were performed on Cray XD2000 at the Center for Computational Astrophysics, National Astronomical Observatory of Japan.
\end{acknowledgements}
\software{XSTAR \citep{Kallman2004}}\clearpage
\bibliography{tanimoto}{}
\bibliographystyle{aasjournalv7}
\end{document}